\documentclass[useAMS,usenatbib]{mn2e}
\usepackage{graphicx}

\newcommand{\ltsima} {$\; \buildrel < \over \sim \;$}
\newcommand{\gtsima} {$\; \buildrel > \over \sim \;$}
\newcommand{\lta} {\lower.5ex\hbox{\ltsima}}
\newcommand{\gta} {\lower.5ex\hbox{\gtsima}}

\newcommand{\lsim}{\raisebox{-.4ex}{$\stackrel{<}{\scriptstyle \sim}$}}
\newcommand{\gsim}{\raisebox{-.4ex}{$\stackrel{>}{\scriptstyle \sim}$}}

\newcommand{\RNum}[1]{\uppercase\expandafter{\romannumeral #1\relax}}

\begin{document}
\title[Newtonian analogue of Kerr black holes]
{Newtonian analogue of corresponding spacetime dynamics of rotating black holes: Implication on black hole accretion. }

\author[Ghosh et al.]
  {Shubhrangshu Ghosh$^{1}$ \thanks{Email address:shubhrang.ghosh@gmail.com}, Tamal Sarkar$^{1,2}$ \thanks{Email address: ta.sa.nbu@hotmail.com}, Arunava Bhadra$^{1}$ \thanks{Email address: aru\_bhadra@yahoo.com}\\ 
   $^{1}$ High Energy and Cosmic Ray Research Centre, University of North Bengal, Post N.B.U, Siliguri 734013, India. \\  
   $^{2}$ University Science Instrumentation Centre,University of North Bengal, Post N.B.U, Siliguri 734013, India. }

\maketitle

\label{firstpage}

\begin{abstract}

Based on the conserved Hamiltonian for a test particle, we have formulated a Newtonian analogue of Kerr spacetime in the `low energy limit of the test particle motion' that, in principle,  can be  comprehensively used to describe general relativistic (GR) features of Kerr spacetime, however, with less accuracy for high spin. The derived potential, which has an explicit velocity dependence, contains the entire relativistic features of corresponding spacetime including the frame dragging effect, unlike other prevailing pseudo-Newtonian potentials (PNPs) for the Kerr metric where such an effect is either totally missing or introduced in a ad hoc manner. The particle dynamics with this potential precisely reproduce the GR results within a maximum $\sim 10 \%$ deviation in energy for a particle orbiting circularly in the vicinity of a rapidly corotating black hole. GR epicyclic frequencies are also well reproduced with the potential, though with a relatively higher percentage of deviation. For counterrotating cases, the obtained potential replicate the GR results with  precise accuracy. The Kerr-Newtonian potential also approximates the radius of marginally stable and marginally bound circular orbits with reasonable accuracy for $a<0.7$. Importantly, the derived potential can imitate the experimentally tested GR effects like perihelion advancement and bending of light with reasonable accuracy. The formulated Kerr-Newtonian potential thus can be useful to study complex accreting plasma dynamics and its implications around rotating BHs in the Newtonian framework, avoiding GR gas dynamical equations.

\end{abstract}

\section{Introduction}

Spinning black holes (BHs) have wide physical implications ranging from exotic frame dragging to controlling some of the highest energetic phenomena in the observed universe. Astrophysical BHs which mostly exhibit in two extreme mass limits: stellar mass BHs of $\sim (5-10) M_{\odot}$ in BH X-ray binaries (BHXRBs) and supermassive BHs (SMBHs) of $\gsim \, 10^6 M_{\odot}$ residing in the center of all galaxies (AGNs and quasars), are realized in the physical universe through the accretion of gaseous plasma around them and its related phenomena (e.g. Bisnovatyi-Kogan \& Lovelace 2001; Ho 2008 and 
references therein). BH spin powers the accretion flow and governs the accretion dynamics, especially in their inner regions 
in the vicinity of it (Meier 1999; Bhattacharya et al. 2010), consequently describing a diverse accretion related phenomena from quasi periodic oscillations (QPOs) (Stella \& Vietri 1999; Mukhopadhyay 2009) to powering astrophysical jets (De Villers et al. 2005; Bhattacharya et al. 2010). Spin of the BH is plausibly responsible for accretion disc precession in its inner regions through Bardeen-Peterson effect (Schawinski et al. 2007) which in turns regulate the precession of relativistic jets.  
Apart from accretion powered jets, astrophysical jets might also be powered by direct extraction of rotational energy of SMBHs in AGNs (Blandford \& Znajek 1977). Recent studies indicate that SMBH spin could enhance the observed luminosity in 
BH accreting systems by several orders in magnitude and might play a predominant role in defining observed AGN classes (Rajesh \& Mukhopadhyay 2010; Mukhopadhyay et al. 2012; Ghosh \& Konar, submitted to MNRAS, and references therein). Galactic mergers drive SMBH binaries to coalescence determining the final state of BH spin (Rezzolla et al. 2008; Mart\'inez-Sansigre \& Rawlings 2011). The co-evolution of SMBHs (both spin and mass) and their host galaxies remains one of the outstanding problems in cosmic
structure formation (Cattaneo et al. 2009). Owing to such an universal and indispensable nature of BH spin, 
its effect on astrophysical processes like accretion related phenomena, could not be possibly ignored. 

BHs are exact classical solutions of field equations in Einstein's theory of general relativity. The formulation of a precise accretion flow model around a central BH requires a combination of numerous physical processes ranging from advective two temperature relativistic plasma dynamics; magnetohydrodynamic turbulent diffusive terms including viscosity, resistivity, thermal conductivity; detailed radiative processes; several local physics and non-linear effects of collisionless plasma (Sharma et al. 2007; Cremaschini et al. 2012). This is a complex and tricky subject, especially accretion flow in the 
vicinity of BHs, where general relativistic (GR) effects are important. It becomes yet more difficult when outflows and jets are included and perturbative affects are incorporated in the flow. Such a complex physical system with GR equations often becomes 
inconceivable in practice. To avoid the GR fluid equations, most of the authors study accretion and its related processes around BHs using fluid equations in the Newtonian framework. Notwithstanding, authors often use simple Newtonian potential without considering the essential GR effects in investigating Keplerian accretion dynamics around nonrotating BHs (Shakura \& Sunayaev 1973; Pringle 1981). The only impression it accommodates from general relativity is that the innermost edge of the disc truncates at the marginally stable circular orbit of Schwarzschild geometry. Use of spherically symmetric Newtonian potential 
often gives satisfactory results for accretion phenomena around non-rotating BHs within a limit of accuracy, excluding the very inner regions of the disc where GR effects are important. However, the exterior solution of rotating BH whose spin is purely a GR effect, is described by Kerr geometry, which without having any spherical symmetry does not have any Newtonian analogue, unlike 
Schwarzschild metric. The only recourse authors employ is pseudo-Newtonian approach by taking into account few relativistic features of Kerr geometry and accommodating it into gas dynamical equations in Newtonian framework, in order to avoid cumbersome GR equations. 

Pseudo-Newtonian potentials (PNPs) have been extensively used in astrophysical literature, especially in regard to accretion flow around BHs after the seminal work of Paczy\'nski \& Witta (1980) (hereinafter PW80). The corresponding potential, although ad hocly introduced, quite precisely reproduce last stable circular orbit in Schwarzschild geometry and has been widely featured in literature to study accretion dynamics around nonrotating BHs. Several other PNPs have been proposed for accretion flows either to describe epicyclic frequency or fluid dynamical aspect around rotating as well as nonrotating BHs in equatorial plane (Nowak \& Wagoner 1991; Artemova et al. 1996; Mukhopadhyay \& Misra 2003, hereinafter MM03). Mukhopadhyay (2002) (hereinafter M02) prescribed a PNP to describe fluid dynamics of accretion flow around a rotating BH in equatorial plane, deriving directly from Kerr metric. Based on this method, Ghosh (2004) developed a PNP corresponding to Hartle-Thorne metric which describes an exterior solution of rotating hard surface. Ghosh and Mukhopadhyay (2007) (hereinafter GM07) formulated a generalized pseudo-Newtonian vector potential useful for studying accretion gas dynamics around a rotating BH in off-equatorial plane. Both these PNPs (M02, GM07) which have been methodologically derived from metric itself are found to be 
valid for entire regime of Kerr parameter, 
however, perturbative effects and epicyclic frequencies are not best described by them. Nonetheless, both the PNPs of M02 have GM07 have been used in several hydrodynamical/hydromagnetic accretion studies with admirable success (Chan et al. 2005; Lipunov \& Gorbovskoy 2007;  Shafee et al. 2008; Bhattacharya et al. 2010). Few other PNPs have also been ad hocly proposed describing generalized Kerr geometry (Semer\'ak \& Karas 1999; Chakrabarti \& Mondal 2006).

Although, PNPs mimic few GR features of corresponding spacetimes to certain extent, however, a single PNP corresponding to a particular metric still lack the uniqueness to describe all GR effects simultaneously, within a reasonable accuracy. Unlike PNPs of M02 and GM07, most of the PNPs are arbitrarily proposed in an ad hoc way without direct correspondence to the metric. PNPs in a generic way are formulated or prescribed to reproduce circular orbits, best suited to study Keplerian accretion flow. Nevertheless, a more fundamental issue regarding PNP is that a PNP is not a physical analogue of local gravity, and is not based on any robust physical theory and does not satisfy Poison equation. PNP is simply a mathematical mimicking of certain GR features of the corresponding metric which is used instead of Newtonian potential in the Newtonian framework fluid equations. Also, certain unique 
GR features like perihelion precession are not well reproduced with most PNPs. Recently, Wegg (2012) ad hocly proposed couple of PNPs by modifying PW80 to reproduce precessional effects in general relativity for orbits with large apoapsis, however, they are not quite effective in the vicinity of the Schwarzschild BH. Things become yet more intriguing in formulating a PNP corresponding to Kerr geometry, as unique features of Kerr spacetime like frame dragging and gravitomagnetic effects necessitate an explicit information of these effects in the corresponding PNP. Although PNPs of M02 and GM07 plausibly contain the information of these effects as they have been derived from the GR metric, however, they do not exhibit them explicitly. Other PNPs simply accommodate these terms in an ad hoc fashion.  

Recently, Tejeda and Rosswog (2013) (hereinafter TR13) formulated a generalized effective potential in the same vein as of Newtonian for a Schwarzschild BH, describing a particle motion around it, based on a proper axiomatic procedure. The potential which is developed directly from the corresponding metric can be viewed as some kind of Newtonian analogue to GR metric which has an explicit dependence on radial velocity and angular velocity of test particle. This generalized potential reproduces exactly several relativistic features of corresponding Schwarzschild geometry. As articulated earlier about 
the importance of rotating BHs in astrophysical scenarios, following TR13, we would pursue to develop a generalized effective potential of a Kerr BH in the equatorial plane for a test particle motion. The potential would then be an appropriate Newtonian or potential analogue of Kerr spacetime which we would refer to as `Kerr-Newtonian' potential. This kind of potential would then be useful to study accretion related phenomena around rotating BHs in a more effective way. 
 
In the next section, we will derive the Kerr-Newtonian potential starting from the Kerr metric. Subsequently in \S 3, we will compare various relativistic features with our potential. In \S 4, we will compare the effectiveness of our potential with 
other existing PNPs in the literature, in reproducing the GR features of Kerr geometry. Finally, we will end up in \S 5 with a discussion and summary.       

\section{Formulation of the generalized potential}

The Kerr spacetime in the Boyer-Lindquist coordinate system is given by  
\begin{eqnarray}
\nonumber
ds^2=-\left(1-\frac{2 \,r_s \, r}{\Sigma}\right)c^2 dt^2-\frac{4ar_s \, r \sin^2 \theta}{\Sigma} c {dt} {d\phi}
+\frac{\Sigma}{\Delta}dr^2\\+\Sigma d\theta^2+\left(r^2+a^2+
\frac{2r_s \, ra^2 \sin^2 \theta}{\Sigma}\right) \sin^2 \theta \, d\phi^2 \, ,
\label{1}
\end{eqnarray}
where $\Delta=r^2+a^2-2r_sr$, $\Sigma=r^2+a^2 \cos^2 \theta$, $r_s= GM/c^2$ and $a=\frac{J}{Mc}$, 
which is called the Kerr parameter. The Lagrangian density of the particle of mass $m$ in the Kerr spacetime is then given by
\begin{eqnarray}
2 {\cal L} =-\left(1-\frac{2 r_s \, r}{\Sigma}\right) c^2 \left(\frac{dt}{d\tau}\right)^2
-\frac{4ar_s \, r \sin^2 \theta \, c}{\Sigma}\frac{dt}{d\tau} \frac{d\phi}{d\tau} \nonumber \\
+\frac{\Sigma}{\Delta} \left(\frac{dr}{d\tau}\right)^2 
+\Sigma \left(\frac{d\theta}{d\tau}\right)^2+\left(r^2+a^2+
\frac{2r_s \, ra^2 \sin^2 \theta}{\Sigma}\right) \nonumber \\
\sin^2\theta \left(\frac{d\phi}{d\tau}\right)^2,
\label{2}
\end{eqnarray}

From the symmetries, we obtain two constants of motion corresponding to two ignorable coordinates $t$ and $\phi$ given by
\begin{eqnarray}
{\cal P}_t =\frac{\partial \cal L}{\partial \tilde{t}} = -\left(1-\frac{2r_s \, r}{\Sigma}\right) c^2 \frac{dt}{d\tau}-
\frac{2ar_s \, r \sin^2 \theta}{\Sigma} c \frac{d\phi}{d\tau} \nonumber \\
={\rm constant}= -\epsilon 
\label{3}
\end{eqnarray}
and 
\begin{eqnarray}
 {\cal P_{\phi}}=\frac{\partial \cal L}{\partial \tilde{\phi}}=-\frac{2ar_s \,r \sin^2 \theta}{\Sigma} c \frac{dt}{d\tau} \nonumber \\ 
+\left(r^2+a^2+\frac{2r_s \, ra^2 \sin^2 \theta}{\Sigma}\right) 
\sin^2\theta \frac{d\phi}{d\tau}={\rm constant} = \lambda \, ,
\label{4}
\end{eqnarray} 

where, $\epsilon$ and $\lambda$ are specific energy and specific angular momentum of the orbiting particle, respectively. Here, $\tilde{t}$ and $\tilde{\phi}$ represent the derivatives of `$t$' and `$\phi$' with respect to proper time $\tau$. 
For particle motion in the equatorial plane $(\theta = {\pi}/2)$, by solving the above two equations we obtain 

\begin{eqnarray}
\frac{dt}{d\tau} =\frac{\frac{\epsilon}{c^2} \left(r^3+a^2 r +2 r_s a^2 \right) 
- \frac{2ar_s \lambda}{c}}{r\Delta} \, ,
\label{5}
\end{eqnarray}
\begin{eqnarray}
\frac{d\phi}{d\tau} =\frac{\frac{\epsilon}{c} 2ar_s + (r - 2r_s) \lambda}{r\Delta}\, .
\label{6}
\end{eqnarray}
Using ${\cal L} = -\frac{1}{2}  m^2 c^2 $ and substituting (\ref{5}) and (\ref{6}) in 
(\ref{2}) we obtain 
\begin{eqnarray}
\frac{\epsilon^2 - c^4}{2c^2} \left(1+\frac{a^2}{r^2}\right)= \frac{1}{2} {\dot r}^2 
\left(\frac{dt}{d\tau}\right)^2 - \frac{r_s a^2}{r^3} \frac{\epsilon^2}{c^2} 
+ \frac{2ar_s \lambda}{r^3} \frac{\epsilon}{c} \nonumber \\ 
- \frac{GM}{r} + \frac{1}{2} \frac{\lambda^2}{r^2} \left(1 - \frac{2 r_s}{r}\right) \, . 
\label{7}
\end{eqnarray}  

Using (\ref{5}) and (\ref{6}) we find 

\begin{eqnarray}
\frac{dt}{d\tau} = \frac{\epsilon}{c^2} \, \frac{r}{\left[(r-2r_s) 
+ \frac{2ar_s}{c} {\dot \phi} \right]} \, .
\label{8}
\end{eqnarray} 

\begin{figure*}
\centering
\includegraphics[width=170mm]{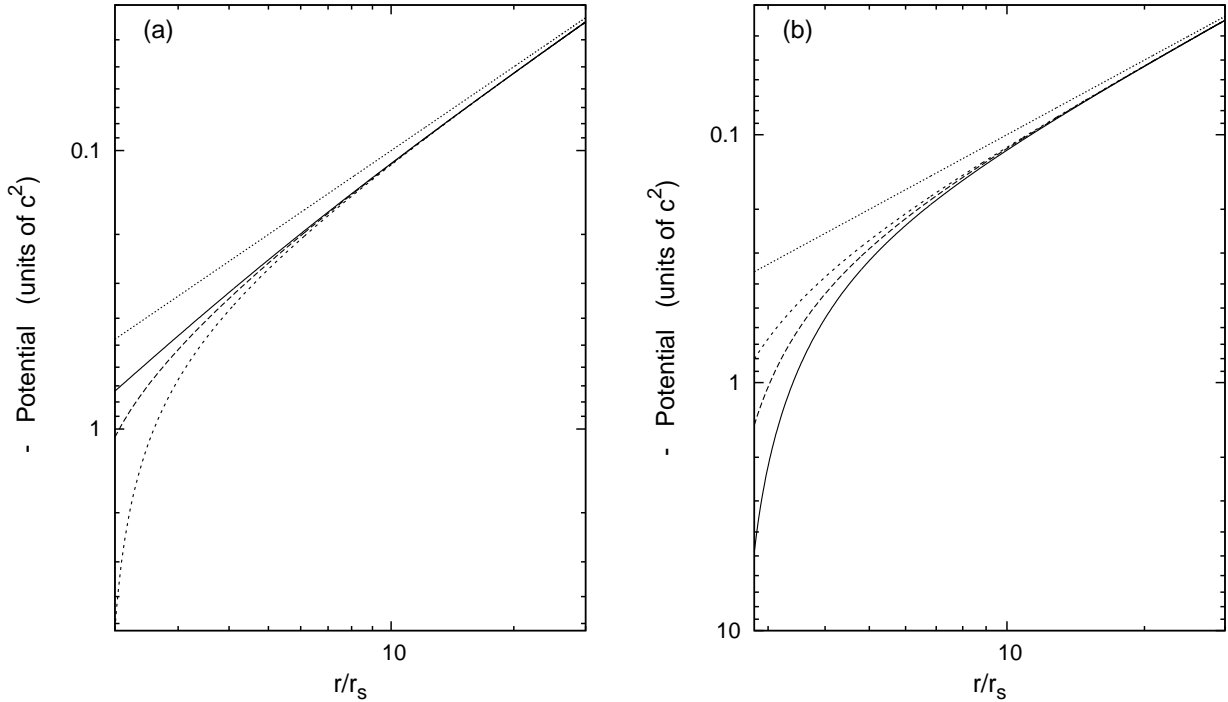}
\caption{Variation of potential with radial distance $r$. Solid, long-dashed, short-dashed and dotted curves in (a) are for Kerr-Newtonian potential with Kerr parameter $a=1$, $a=0.5$, Schwarzschild-Newtonian potential ($a=0$) and Newtonian potential respectively. Similarly Solid, long-dashed,  short-dashed and dotted curves in (b) are for Kerr parameter $a=-1$, 
$a=-0.5$, Schwarzschild case and Newtonian case respectively. Potential in y-axis has negative values expressed in units of $c^2$. 
$r$ and $a$ are expressed in units of $r_s$. Both x-axis and y-axis are in logarithmic scale.  } 
\label{Fig1}
\end{figure*}

The basis of our potential formulation is the low energy limit of the test particle motion (TR13), which is $\epsilon/c^2 \sim 1$. We write $E = \frac{\epsilon^2 - c^4}{2c^2}$ considering a locally inertial frame for test particle motion which will reduce to the total 
mechanical energy ($\equiv$ Hamiltonian) in Newtonian mechanics in nonrelativistic limit with $a=0$. Second term in the above definition of $E$ is the rest mass energy of the particle which is subtracted from relativistic energy owing to the low energy limit, in analogy to Newtonian Hamiltonian. Computing $\lambda$ from (\ref{6}) and substituting in (\ref{7}) and using (\ref{8}), we finally obtain the generalized Hamiltonian ($E_{\rm GK}$) of test particle around Kerr spacetime in low energy limit as 

\begin{eqnarray} 
E_{\rm GK}= -\frac{GM}{r} + \left(\frac{1}{2} {\dot r}^2  \frac{r-2r_s}{\Delta} 
+ \frac{\Delta}{2r} {\dot \phi}^2 \right) \, \frac{r^3}{\left[(r-2r_s) 
+ \frac{2ar_s}{c} {\dot \phi} \right]^2} \, , 
\label{9}
\end{eqnarray}

where, overdots represent the derivative with respect to coordinate time $t$. With $a=0$, $E_{\rm GK}$ reduces to that of Schwarzschild geometry. The generalized Hamiltonian $E_{\rm GK}$ in the low energy limit should be equivalent to the Hamiltonian in the Newtonian framework. The effective Hamiltonian in the Newtonian regime with the generalized potential in the equatorial plane will then be equivalent to $E_{\rm GK}$ in (\ref{9}). Thus 

\begin{eqnarray}
E_{\rm GK} \equiv  \frac{1}{2} \left({\dot r}^2 + r^2 {\dot \phi}^2 \right) 
+ V_{\rm GK} - {\dot r} \frac{\partial V_{\rm GK}}{\partial {\dot r}} 
- {\dot \phi} \frac{\partial V_{\rm GK}}{\partial {\dot \phi}} \, , 
\label{10}
\end{eqnarray}

where, $T = \frac{1}{2}({\dot r}^2 + r^2 {\dot \phi}^2)$ is the nonrelativistic specific kinetic energy of the test particle. $V_{\rm GK}$ is the most generalized form of the potential in Newtonian analogue of Kerr spacetime in the spherical geometry with test particle motion 
in the equatorial plane, which contains the entire information of the source. The potential $V_{\rm GK}$ is then given by 

\begin{eqnarray}
V_{\rm GK}  = -\frac{GM}{r} (1 - \omega \dot \phi) - \frac{\left( {\mathcal G}_{1} {\dot r}^2 + {\mathcal G}_{2} 
r^2 {\dot \phi}^2 \right)}{2 \left(1+\omega {\dot \phi} \right)} +  \frac{{\dot r}^2 + r^2 {\dot \phi}^2}{2} \, , 
\label{11}
\end{eqnarray}

where 

\begin{eqnarray}
{\mathcal G}_{1} = \frac{r^3}{(r-2r_s)\Delta} , \, \, \,   
{\mathcal G}_{2} = \frac{\Delta}{(r-2r_s)^2} \, .
\label{12}
\end{eqnarray}

Note that all the dynamical quantities expressed are specific quantities. In the Newtonian limit ${\mathcal G}_{1} = {\mathcal G}_{2} = 1$. $\omega = {2ar_s}/{c(r-2r_s)}$. $\omega \dot \phi$ in the potential in (\ref{11}) arises due to the effect of frame dragging. Potential $V_{\rm GK} (\equiv V_{\rm KN})$ is a modified potential deviating from exact Newtonian (spherical symmetric part). `${\rm KN}$' symbolizes `Kerr-Newtonian'. The potential is an explicit velocity dependent potential
containing all gravitational effects of Kerr spacetime for a stationary observer. Thus, the potential in (\ref{11}) contains 
the explicit information of gravitomagnetic and frame dragging effects which has been obtained directly from the Kerr metric by solving geodesic equations 
of motion. Putting $a=0$, the potential reduces to that in Schwarzschild geometry. Unlike most other PNPs which are either derived or prescribed for particle motion in circular orbit, the potential in (\ref{11}) is applicable for generalized orbital dynamics. It is to be noted that we have restricted ourselves in deriving a Kerr-Newtonian potential corresponding to a particle motion in the equatorial plane. Formulation of a more generalized Kerr-Newtonian potential for off-equatorial particle orbits is immensely complicated within our present approach, where the necessary use of Carter constant seems to be a prerequisite (see GM07). Such a study would be pursued in the near future. 

Although the Kerr-Newtonian potential, in principle, should precisely reproduce all orbits in exact Kerr geometry, the form of the potential in (\ref{11}) gets diverge at $r = 2r_s = 2GM/c^2$. This is precisely happening owing to the presence of the $\left[(r-2r_s) + \frac{2ar_s}{c} {\dot \phi} \right]^2$ in the denominator of Hamiltonian $E_{\rm GK}$ in (\ref{9}), which has been obtained 
while replacing the conserved specific angular momentum $\lambda$ by ${\dot \phi}$. Thus, the potential in the form given in (\ref{11}) would not be useful within the range $r \, \lsim \, 2r_s$. Note that for Kerr BH, the horizon $r_H = r_s$ for maximal spin. However, such a radial zone of range $r \, \lsim \, 2 \, r_s$ is in the extreme vicinity of the rotating BH, which either lies within the 
ergosphere for a certain range of $a$ or having a direct ergospheric effect. Moreover, at $r \, \lsim \, 2 \, r_s$, 
the notion of potential indeed becomes insignificant and exact GR equations become relevant, where ergospheric effects would dominate. The accretion powered phenomena which we would be more interested in are more relevant at much outer radii, as most of the observed phenomena related to BH accretion occur at radii much beyond $\sim 2 \, r_s$. Also, it is to be noted that for lesser BH spin, 
$r_H$ is much greater than $r_s$ for which the inner accretion edge is way beyond $\sim 2 \, r_s$.   

In Fig. 1, we show the variation of the Kerr–Newtonian potential
with r for both prograde and retrograde circular orbits and we
compare them with the Schwarzschild and Newtonian cases. It is being seen that for corotating case (Fig. 1a), the magnitude of the corresponding Kerr-Newtonian potential is less than that with respect to Schwarschild spacetime in the inner regions 
of the central BH, and decreases with the increase in Kerr parameter $a$. 
This occurs exactly due to the effect of frame dragging. With the increase in 
$a$, the effect of frame dragging increases which tends to diminish the 
radial effect of Kerr-Newtonian potential. This property of Kerr spacetime 
has a direct consequence on the accreting plasma in the vicinity of rotating BHs, by 
providing an additional boost to propel matter and radiation out of the 
accretion flow. On the contrary, for counterrotating particle orbits, the magnitude of Kerr-Newtonian potential is much higher as compared to that in Schwarzschild geometry, which increases with the increase in $a$ (Fig. 1b) 

The Lagrangian of the particle in the presence of this Kerr-Newtonian potential is given by 

\begin{eqnarray}
{\cal L}_{\rm KN} = \frac{GM}{r} (1 - \omega {\dot \phi}) + \frac{\left( {\mathcal G}_{1} {\dot r}^2 + {\mathcal G}_{2} 
r^2 {\dot \phi}^2 \right)}{2\left(1+\omega {\dot \phi} \right) }  \, ,
\label{13}
\end{eqnarray}

which exactly reduces to that in Schwarzschild geometry with $a=0$. Specific angular momentum which is a constant of motion corresponding to Kerr-Newtonian potential is then given by 

\begin{eqnarray}
\lambda_{\rm KN} = \frac{\partial {\cal L}_{\rm KN}}{\partial {\dot \phi}} = - \frac{GM\omega}{r} + 
\frac{{\mathcal G}_{2} r^2 {\dot \phi} \left(2 + \omega {\dot \phi} \right)}{2 \left(1+ \omega {\dot \phi} \right)^2}
- \frac{{\mathcal G}_{1} {\dot r}^2 \omega }{2 \left(1+ \omega {\dot \phi} \right)^2}  \, . 
\label{14}
\end{eqnarray}

Obtaining the specific Hamiltonian from (\ref{13}), the radial motion of the particle in the presence of this 
potential is then given by 

\begin{eqnarray}
{\dot r}^2 = \frac{2}{{\mathcal G}_{1}} \left(E_{\rm KN} + \frac{GM}{r} \right) \left(1+ \omega {\dot \phi} \right)^2
- \frac{{\mathcal G}_{2}}{{\mathcal G}_{1}} r^2 {\dot \phi}^2  \, .
\label{15}
\end{eqnarray}

$E_{\rm KN}$ is the conserved specific Hamiltonian of the particle motion in Kerr-Newtonian which is equivalent to $E_{\rm GK}$. ${\dot r}$ is identical to the expression in exact Kerr geometry in low energy limit. Next we compute the equations of motion of test particle using the Kerr-Newtonian potential. For $r$ coordinate we obtain 

\begin{eqnarray}
\left(1-\frac{{\mathcal{A}}_3}{{\mathcal{B}}_1} {\mathcal{B}}_5 {\dot r}^2 \right) \ddot{r} +  \left[{\mathcal{A}}_1 + \frac{{\mathcal{A}}_3}{{\mathcal{B}}_1} \left({\mathcal{B}}_2 +  {\mathcal{B}}_3  + {\mathcal{B}}_4   \right) \right] {\dot r}^2 \nonumber \\
- {\mathcal{A}}_2 {\dot \phi}^2 + {\mathcal{A}}_4 
+ \frac{{\mathcal{A}}_3}{{\mathcal{B}}_1} {\mathcal{B}}_6  {\dot r}^4 \, = \, 0 \, .
\label{16}
\end{eqnarray}

Similarly for $\phi$ coordinate we have 

\begin{eqnarray}
\left(1-\frac{{\mathcal{A}}_3}{{\mathcal{B}}_1} {\mathcal{B}}_5 {\dot r}^2 \right) \ddot{\phi} + \left[\frac{1}{{\mathcal{B}}_1} \left({\mathcal{B}}_2  +  {\mathcal{B}}_3 + {\mathcal{B}}_4  + {\mathcal{A}}_4  {\mathcal{B}}_5 \right)  \right]  {\dot r} \nonumber \\
- \frac{{\mathcal{A}}_2 {\mathcal{B}}_5}{{\mathcal{B}}_1} {\dot r} {\dot \phi}^2 
+ \frac{1}{{\mathcal{B}}_1} \left({\mathcal{B}}_6  + {\mathcal{A}}_1 {\mathcal{B}}_5 \right)  {\dot r}^3 \, = \, 0 \, .
\label{17}
\end{eqnarray}

Here, \\
$$
{\mathcal{A}}_1 = \frac{1}{2(r-2r_s)} 
\left[\frac{2 a^2 (r-3r_s) - 4 r r_s (r-2r_s)}{r\Delta} + \frac{\omega {\dot \phi}}{1 + \omega {\dot \phi}} \right ]  , 
$$
$$
{\mathcal{A}}_2 = \frac{{\mathcal G}_{2}}{2r} \left[ 2 (r-3r_s)(r-2r_s) - 4 \frac{r_s}{r} a^2 \, + \, \Delta \frac{\omega {\dot \phi} }{1+ \omega {\dot \phi}}   \right] , 
$$
$$
{\mathcal{A}}_3 = \frac{\omega}{1 + \omega {\dot \phi}} , \, \, 
{\mathcal{A}}_4 = \frac{GM}{r^2 {\mathcal G}_{1}} \left[1 - \frac{4ar_s}{c} \frac{r-r_s}{(r-2r_s)^2} {\dot \phi} \right] ,
$$
$$
{\mathcal{B}}_1 = \left( {\mathcal G}_2 r^2  +  {\mathcal G}_1 \omega^2 {\dot r}^2 \right) , \, \,  
{\mathcal{B}}_2 =  \frac{r^2 \dot \phi}{2(r-2r_s)} {\mathcal G}_2 \, \omega {\dot \phi} \left(3 +  \omega {\dot \phi} \right)  
$$
$$
{\mathcal{B}}_3 = \frac{r^2 \dot \phi \left(1+   \omega {\dot \phi}\right)}{2(r-2r_s)} \left[\frac{2 (r-3r_s)(r-2r_s) - 4 \frac{r_s}{r} a^2 }{(r-2r_s)^2} 
\left(2+ \omega {\dot \phi}\right) \right] , 
$$ 
$$
{\mathcal{B}}_4 = \frac{4 G M r_s a (r-r_s) \left(1+\omega {\dot \phi} \right)^3}{c r^2 (r-2 r_s)^2 } , \, \,  {\mathcal{B}}_5 = {\omega} {\mathcal G}_1 \left(1+\omega {\dot \phi} \right) 
$$ 
and 
$$
 {\mathcal{B}}_6 = \frac{\omega}{2} \left[ \frac{{\mathcal G}_1 \left(1-\omega {\dot \phi} \right)}{r-2r_s} 
- \frac{ 2 a^2 r^2 (r-3r_s) - 4 r^3 r_s (r-2r_s) }{(r-2r_s)^2 \frac{\Delta^2}{\left(1+  \omega {\dot \phi} \right)}} \right]\, .
$$

The equations (\ref{14}), (\ref{15}), (\ref{16}) and (\ref{17}) will provide a complete particle dynamics around Kerr BHs in 
Kerr-Newtonian framework. They reduce to the expressions corresponding to Schwarzschild case, with $a=0$.

\section{Comparison of GR features with the Kerr-Newtonian potential}

In this section, we will compare various GR features with the Kerr-Newtonian potential for different values of Kerr parameter $a$. As argued earlier, we will use the form of potential given in (\ref{11}) in which case the potential will be generically valid beyond $r \, \gsim \, 2r_s$. 

\subsection{Dynamics of circular orbit} 

Circular orbit of the test particle is determined by conditions 
\begin{eqnarray}
{\dot r} = 0,  \, \, \,  \ddot{r} = 0 \, .
\label{18}
\end{eqnarray}
We use the said conditions for circular orbits using (\ref{14}), (\ref{15}) and (\ref{16}) from 
where we obtain specific angular momentum $\lambda_{\rm KN} {\vert}_C$, specific Hamiltonian $E_{\rm KN} {\vert}_C$ and 
specific angular velocity ${\dot \phi}_{\rm KN} {\vert}_C$ numerically. 
The symbol `${\vert}_C$' corresponds to the dynamical 
quantities in circular orbit. Alternatively, $\lambda_{\rm KN} {\vert}_C$ and $E_{\rm KN} {\vert}_C$ can be directly obtained from (\ref{15}) by replacing 
${\dot \phi}$ with $\lambda_{\rm KN}$ and its corresponding derivative, and subsequently using prerequisite circular orbit conditions. In that case it is then 
possible for us to obtain analytical expressions for $\lambda_{\rm KN} {\vert}_C$ and $E_{\rm KN} {\vert}_C$, respectively. $\lambda_{\rm KN} {\vert}_C$ is then given by
\begin{eqnarray}
\lambda_{\rm KN} {\vert}_C=\frac{-{\mathcal{Q}}_1 \pm \sqrt{{\mathcal{Q}}^2_1 
-4  {\mathcal{R}}_1 }}{2 } \, , 
\label{19}
\end{eqnarray}
where, 
$$
{\mathcal{Q}}_1 = \left[ \frac{4a^3 r r_s \, c- 6r_s \, a\, c\, r\left(r^2+a^2\right)}
{a^2 r \left(r-2r_s \right)-r\left(r-3r_s\right)\left(r^2 + a^2 \right) } \right] , 
$$
$$
{\mathcal{R}}_1 = \left[ \frac{GM \left(r^2+ a^2 \right) \left[ r \left(r^2 + 3a^2 \right)-2a^2 r \right]}
{a^2 r \left(r-2r_s \right)-r\left(r-3r_s\right)\left(r^2 + a^2 \right) } \right] \, . 
$$
Similarly, $E_{\rm KN} {\vert}_C$ is given by 
\begin{eqnarray}
E_{\rm KN} {\vert}_C = \frac{\frac{\lambda^2_{\rm KN} {\vert}_C }{2} 
\left(r-2r_s \right) - GMr^{2}}{r^{3}\left(1+ \frac{a^2}{r^2} \right)} \nonumber \\
+ \frac{r_s \left(2 \, a \, c \, \lambda_{\rm KN}{\vert}_C - a^2 c^2 \right)}{r^{3}\left(1+ \frac{a^2}{r^2} \right)}
\label{20}
\end{eqnarray}
${\dot \phi}_{\rm KN} {\vert}_C$ is then computed from a quadratic relation ${\mathcal{P}}_2 \, {\dot \phi}^2_{\rm KN} {\vert}_C 
+ {\mathcal{Q}}_2 \, {\dot \phi}_{\rm KN} {\vert}_C + {\mathcal{R}}_2 = 0 $ obtained using (\ref{14}), where, 
$$
{\mathcal{P}}_2= \left[\omega^{2} \left(\lambda_{\rm KN} {\vert}_C+\frac{GM\omega}{r} \right)-\frac{{\mathcal{G}}_2 r^{2}\omega}{2}\right] , 
$$
$$ 
{\mathcal{Q}}_2 = \left[2\omega \left(\lambda_{\rm KN} {\vert}_C + \frac{GM}{r} \right)-{\mathcal{G}}_2 r^{2} \right] , \, \, {\mathcal{R}}_2 = \left(\lambda_{\rm KN} {\vert}_C+\frac{GM \omega}{r} \right) . 
$$
For $a=0$, this quadratic relation becomes linear and reduces to that in TR13. For $a \not= 0$, we obtain the physically correct solution of ${\dot \phi}_{\rm KN} {\vert}_C$, given by 
\begin{eqnarray}
{\Omega}_{\rm KN} {\vert}_C \equiv {\dot \phi}_{\rm KN} {\vert}_C = \frac{-{\mathcal{Q}}_2-\sqrt{{\mathcal{Q}}^2_2 
-4 {\mathcal{P}}_2 {\mathcal{R}}_2 }}{2 {\mathcal{P}}_2} \, , 
\label{21}
\end{eqnarray}
The corresponding relativistic results in Kerr geometry are given by (Bardeen 1973) 
\begin{eqnarray}
 \lambda_{K} {\vert}_C = \frac{\sqrt{GMr} \left(r^2 - 2 a \sqrt{r_s r} + a^2 \right)}
{r \left(r^2 - 3 r_s r + 2 a \sqrt{r_s r} \right)^{1/2}} \, ,
\label{22}
\end{eqnarray}
\begin{eqnarray}
 \frac{\epsilon_{K} {\vert}_C}{c^2} =  \frac{\left(r^2 - 2 {r_s r} + a \sqrt{r_s r} \right)}
{r \, \left(r^2 - 3 r_s r + 2 a \sqrt{r_s r} \right)^{1/2}} \, ,
\label{23}
\end{eqnarray}
\begin{eqnarray}
\Omega_{K} {\vert}_C \equiv {\dot \phi}_{K} {\vert}_C = \frac{\sqrt{GM}}{r^{3/2} + a r^{1/2}_s} \, .
\label{24}
\end{eqnarray} 
Note that the actual specific Hamiltonian in Kerr geometry is 
$E_K {\vert}_C = \frac{\epsilon^2_{K} {\vert}_C - c^4}{2c^2} $, which is 
being actually plotted. 

\newcounter{tempequationcounter}
\begin{figure*}
\normalsize
\setcounter{tempequationcounter}{\value{equation}}
\begin{eqnarray}  
\setcounter{equation}{28}
\kappa^2   
= - \, {{\dot \phi} {\vert}_C}^2 \left[1 - a^2 (r^2 - 10 r r_s + 10 r^2_s) + 8 \frac{r_s (r-r_s) a^4}{r^3 (r-2 r_s)^3} \right]
- \, {{\dot \phi} {\vert}_C}^2   
\left[\frac{\Delta}{2 r (r - r_s)} -\frac{\Delta (3 r - 2 r_s) a^2}{2 r^2 (r - 2 r_s)^3} \right] \frac{\omega \, {\dot \phi} {\vert}_C}{1 + \omega \, {\dot \phi} {\vert}_C} \nonumber \\
+ \, \frac{{{\dot \phi} {\vert}_C}^2}{r-2r_s} 
\left[ {\mathcal D}_1  \frac{\omega \, {\dot \phi} {\vert}_C}{1 + \omega \, {\dot \phi} {\vert}_C} + {\mathcal D}_2  \frac{\omega \, {\dot \phi} {\vert}_C \, \left(1-\omega \, {\dot \phi} {\vert}_C \right)}{\left(1 + \omega \ {\dot \phi} {\vert}_C\right)^2} \right] \, - \frac{2GM}{r^5} 
\left[(r-2r_s)(r-4r_s) 
+ a^2 \left(2-\frac{5 r_s}{r} \right) \right] \left(1 + \omega \, {\dot \phi} {\vert}_C \right) \nonumber \\
+  \frac{4 G M r_s}{r^5} \frac{a}{c} \frac{r-r_s}{(r-2 r_s)^2} +  {\mathcal F}_1 {\mathcal F}_2 
\left[\Delta \frac{5 r^2 - 14 r r_s + 10 r^2_s}{r(r-r_s)} - 2 (r-r_s) (r-2r_s) \right] {\dot \phi} {\vert}_C \, \left(1 + \omega \ {\dot \phi} {\vert}_C \right)  
\label{28}
\end{eqnarray}
\setcounter{equation}{\value{tempequationcounter}}  
\hrulefill
\vspace*{4pt}
\end{figure*}

Figure 2 shows the variation of specific angular momentum with $r$ for both corotating and counterrotating circular orbits with the Kerr-Newtonian potential and has been compared with the corresponding relativistic geometry. The angular momentum profiles corresponding to the Kerr-Newtonian potential reproduce the GR results quite accurately. 

\begin{figure*}[h]
\centering
\includegraphics[width=130mm]{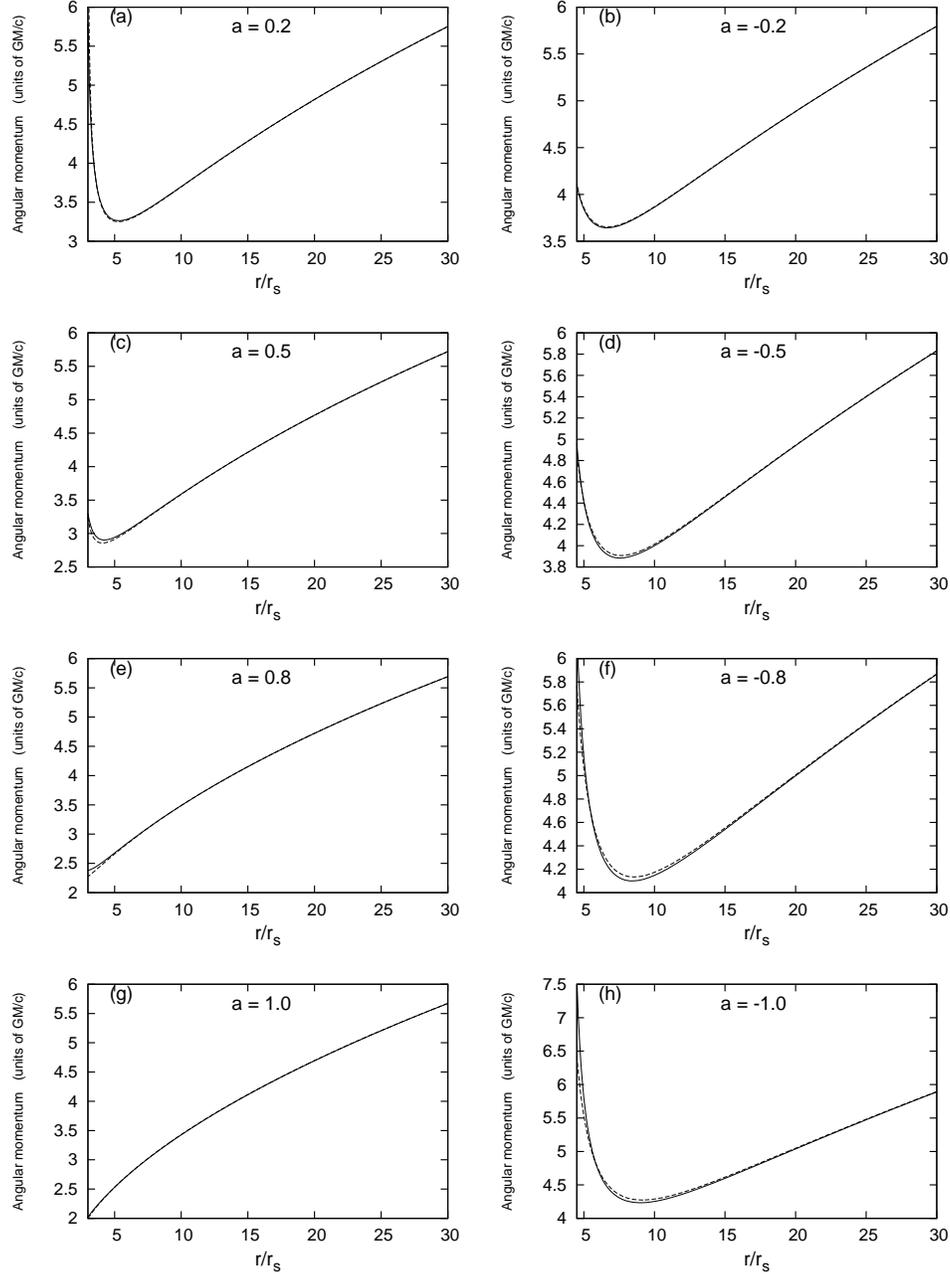}
\caption{Variation of specific angular momentum with radial distance $r$ for 
both corotating and counterrotating circular orbits. Solid and dashed curves are for exact Kerr geometry and Kerr-Newtonian framework respectively.
$r$ and $a$ are expressed in units of $r_s$. Specific angular momentum is 
in units of $GM/c$
 }
\label{Fig2}
\end{figure*}

In Fig. 3 we exhibit similar profiles for corresponding specific Hamiltonian with the Kerr-Newtonian potential, which lies within an error of $\sim 10 \%$, in the vicinity of rapidly corotating BH. However, the counterrotating GR results are reproduced with the Kerr-Newtonian potential with precise accuracy. The angular frequency profiles are displayed in Fig. 4, which too reproduce the GR results, however, with less accuracy in the inner regions of the flow for high BH spin. A maximum error of $\sim 36 \%$ is obtained in the vicinity of an extremely corotating BH. Conversely, for counterrotating particle orbits, the Kerr-Newtonian potential quite accurately reproduces the corresponding GR values.   

\begin{figure*}[h]
\centering
\includegraphics[width=170mm]{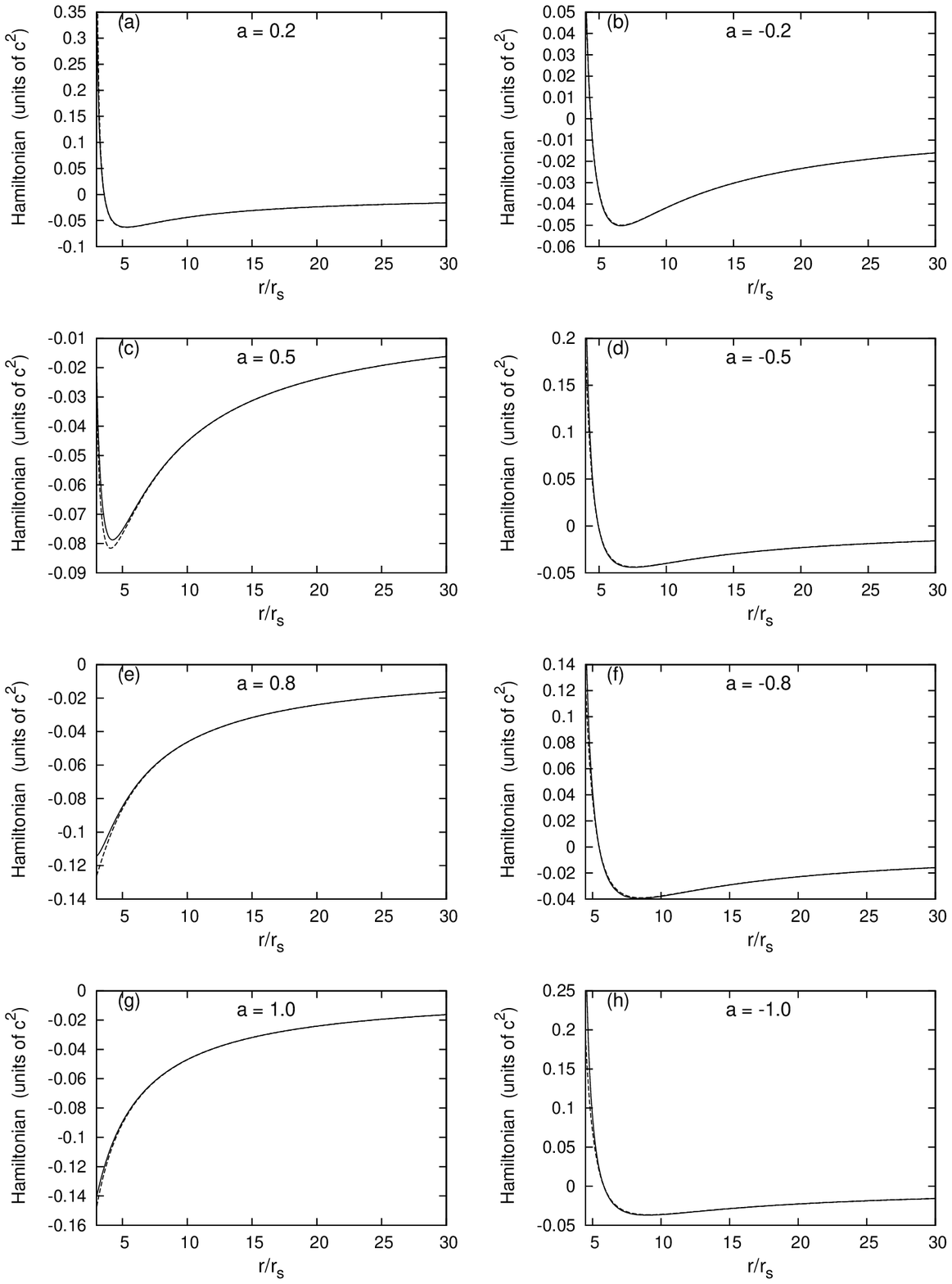}
\caption{Similar to that in Fig. 2, but variation of specific Hamiltonian of particle in circular orbit with radial distance $r$, in units of $c^2$. Other parameters are same as in Fig. 2
 }
\label{Fig3}
\end{figure*}

\begin{figure*}[h]
\centering
\includegraphics[width=170mm]{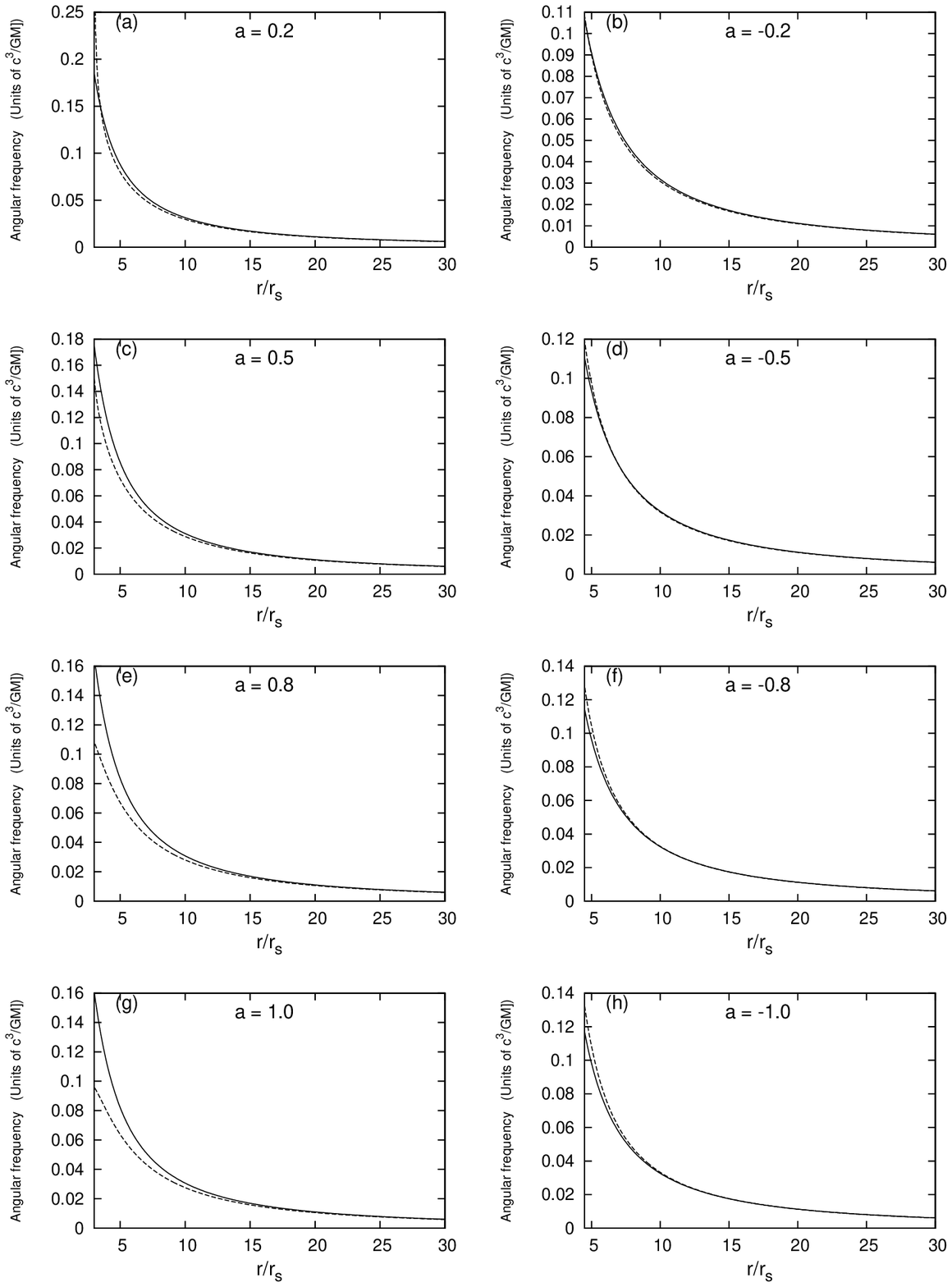}
\caption{Same as that in figures 2, 3, but variation of angular frequency  
of particle in circular orbit with $r$, in units of $c^3/{GM}$.
 }
\label{Fig4}
\end{figure*}

The two salient GR features corresponding to the particle motion in the circular orbit in Kerr geometry are marginally stable $\left(r_{\rm ms} \right)$ 
and marginally bound orbit $\left(r_{\rm mb} \right)$. As usual, we use the conditions $\frac{d\lambda_{\rm KN} {\vert}_C}{dr} = 0$ and $E_{\rm KN} {\vert}_C = 0$ using (\ref{19}) and (\ref{20}) to obtain numerical values of $r_{\rm ms}$ and 
$r_{\rm mb}$ respectively, corresponding to the Kerr-Newtonian potential. It is found that for counterrotating case, $r_{\rm mb}$ is almost exactly replicated with the Kerr-Newtonian potential and $r_{\rm ms}$ is precisely reproduced within an maximum error of $\sim 1.2 \%$, as depicted in Fig. 5. However, for corotating case, we obtain real solution for $r_{\rm ms}$ and $r_{\rm mb}$ up to Kerr parameter $a \sim 0.7$. $r_{\rm mb} $ is being exactly reproduced, however $r_{\rm ms}$ is being reproduced within a reasonable accuracy with a maximum error margin of $\sim 10 \%$, up to the specified value of $a \sim 0.7$ (see Fig. 5). This limiting description of $r_{\rm ms} $ and $r_{\rm mb} $ up to $a \sim 0.7$ is due to the expression under square root in (\ref{19}) which for $a > 0.7$ becomes negative at a radial distance larger than $r_{\rm ms} $ and $r_{\rm mb} $. Note that, circular geodesics could still be described for $a \, \gsim \, 0.7$, however, at radii $r \, \gsim \, 3 \, r_s$. 

\subsection{Orbital perturbation}

Perturbation in accretion flow is mostly studied, in understanding 
the instabilities in the accreting system. Small perturbation in the flow which is linked to the epicyclic 
frequency, and its coupling to BH spin, can be related to QPOs in BHXRBs. TR13 computed the epicyclic 
frequency for a test particle motion with their Schwarzschild-Newtonian 
potential, and compared with the exact GR value, which they found to be highly 
accurate. Using (\ref{16}) and 
(\ref{17}), we then estimate the radial epicyclic frequency for a test particle motion in circular orbit in the equatorial plane, which 
will be influenced by the spin of BH. $r$ and $\phi$ and their derivatives will be perturbed according to 
\begin{eqnarray}
r \to r + \delta r,  \, \, \, \dot r \to \delta {\dot r},  \, \, \, \ddot{r} \to \delta \ddot{r} \, , 
\label{25}
\end{eqnarray}
\begin{eqnarray}
\phi \to \phi + \delta \phi, \, \, \,  {\dot \phi} \to {\dot \phi} {\vert}_C + \delta {\dot \phi}, \, \, \,  \ddot{\phi} = \delta {\ddot{\phi}} \, .
\label{26}
\end{eqnarray}
Inserting equations (\ref{25}) and (\ref{26}) into equations (\ref{16}) and 
(\ref{17}), and using equation (\ref{14}), we obtain the linearized 
perturbed equations. By solving these, the radial epicyclic frequency $\kappa$ is 
computed, given by (\ref{28}). ${\mathcal D}_1, {\mathcal D}_2, {\mathcal F}_1$ 
and ${\mathcal F}_2$ in (\ref{28}) are given in appendix 2. The expression for $\kappa$ in (\ref{28}) exactly reduces to that in Schwarzschild case with $a=0$. Due to cumbersome and long nature of equations, the derivation of $\kappa$ is furnished in appendix 2. We compare the value of $\kappa$ corresponding to  Kerr-Newtonian potential with the exact relativistic result in Kerr geometry, which is given by (Semer\'ak \& Z\'acek 2000)
\begin{eqnarray}  
{\kappa {\vert}_K}^2 = \left(\frac{\Omega_{K} {\vert}_C}{r} \right)^2 
\left[\Delta - 4 (\sqrt(r_s r) - a)^2 \right] \, .
\label{27}
\end{eqnarray}
It needs to be mentioned that we have only derived the radial epicyclic 
frequency with no expression for vertical epicyclic frequency. It is owing to the fact that to have an expression for vertical epicyclic frequency or to 
study perturbations perpendicular to the equatorial plane, one needs to derive a 
`$\theta$' dependent Kerr-Newtonian potential valid for off-equatorial orbital trajectory, and subsequently, a `$\theta$' dependent equation of motion. This, however, is beyond the scope of our present study. 

Figure 6 shows the comparison of radial dependence of $\kappa$ obtained from the Kerr-Newtonian potential with that in general relativity, which  
exhibits a maximum error of $\sim 35 \%$ in the vicinity of a rapidly 
corotating BH. Here too, the Kerr-Newtonian potential quite precisely reproduce the corresponding GR results for counterrotating BH.

\subsection{Orbital precession} 

Using (\ref{14}) and (\ref{15}), we compute ${d\phi}/{dr}$ in 
Kerr-Newtonian which we compare with the corresponding GR expression obtained 
using (\ref{6}), (\ref{7}) and (\ref{8}). In Schwarzschild case, the expressions 
of ${d\phi}/{dr}$ in both Schwarzschild-Newtonian as well as in exact general relativity are 
similar, giving identical orbital trajectory and perihelion precession (TR13). Owing to this exactness, it guarantees that the bending of light or gravitational lensing 
(Bhadra et al. 2010) in Schwarzschild-Newtonian framework would also reproduce identical GR result. Nevertheless, in Fig. 7, we compare ${d\phi}/{dr}$ as a function of $r$ corresponding to both Kerr-Newtonian potential and its GR counterpart, in the low energy limit of the test particle motion. It shows that ${d\phi}/{dr}$ corresponding to Kerr-Newtonian potential 
is almost identical with the corresponding GR result. In Fig. 8, we display elliptic like trajectories for 
particle orbit corresponding to Kerr-Newtonian potential in the $x-y$ plane, obtained from the equations of motion, and compare the nature of trajectories with the GR results for corotating BH. We obtain the plots of elliptical trajectory using Cartesian transformation adopting the method of Euler-Cromer algorithm, which preserves energy conservation. For all cases the test particle starts from an apoapsis $r_a = 60 \, r_s$ with a fixed eccentricity $e = 0.714$. Figure 8 shows that the orbital trajectory corresponding to Kerr-Newtonian potential resembles the GR result well, however with less accuracy for rapidly spinning BH. The value of the apsidal precession can be estimated using the relation of orbital trajectory. The apsidal precession or the 
perihelion advancement $\Psi$ is given by the relation 
\begin{eqnarray}
\setcounter{equation}{29}
\Psi = \Pi - \pi \equiv \int_{r_p}^{r_a} \frac{d \phi}{dr} \, dr - \pi  \, ,
\label{29}
\end{eqnarray}
where $\Pi$  is the usual half orbital period of the test particle and $r_p$ is the periapsis of the orbit. Alternatively, we can easily compute the apsidal precession directly from the trajectory profiles. In Table 1, we display the values of apsidal precession corresponding to Kerr-Newtonian potential and compare them with the GR results, for different values of Kerr parameter $a$. We use similar set of orbital parameters as used in Fig. 8. It is being found that the maximum deviation of $\Psi$ for Kerr-Newtonian potential
from that of the exact GR result is not more that $\sim 12 \%$ \footnote
{${\rm Error(percent)} = \frac{\vert {\rm GR}_{\rm values} - {\rm KN}_{\rm values} \vert}{{\rm GR}_{\rm values}} \times 100$ }
(fourth column of Table 1), corresponding to extremally rotating Kerr BH. 

\begin{figure}
\includegraphics[width=85mm]{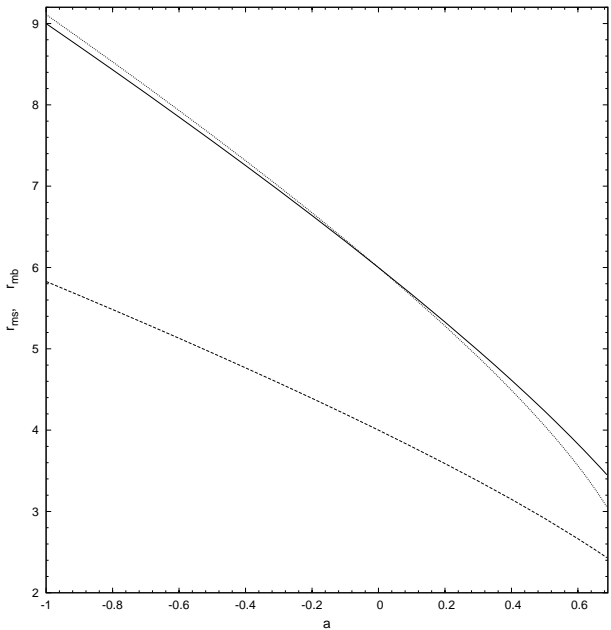}
\caption{Locations of marginally stable $\left(r_{\rm ms} \right)$ and marginally bound orbit $\left(r_{\rm mb}\right)$ for both corotating and counterrotating Kerr BHs. Solid and short-dashed lines are for marginally stable orbits corresponding to Kerr geometry and Kerr-Newtonian potential, respectively. Long-dashed curve is for marginally bound orbit corresponding to the Kerr-Newtonian potential which coincides with the GR result. 
 }
\label{Fig5}
\end{figure}

\begin{figure*}
\centering
\includegraphics[width=140mm]{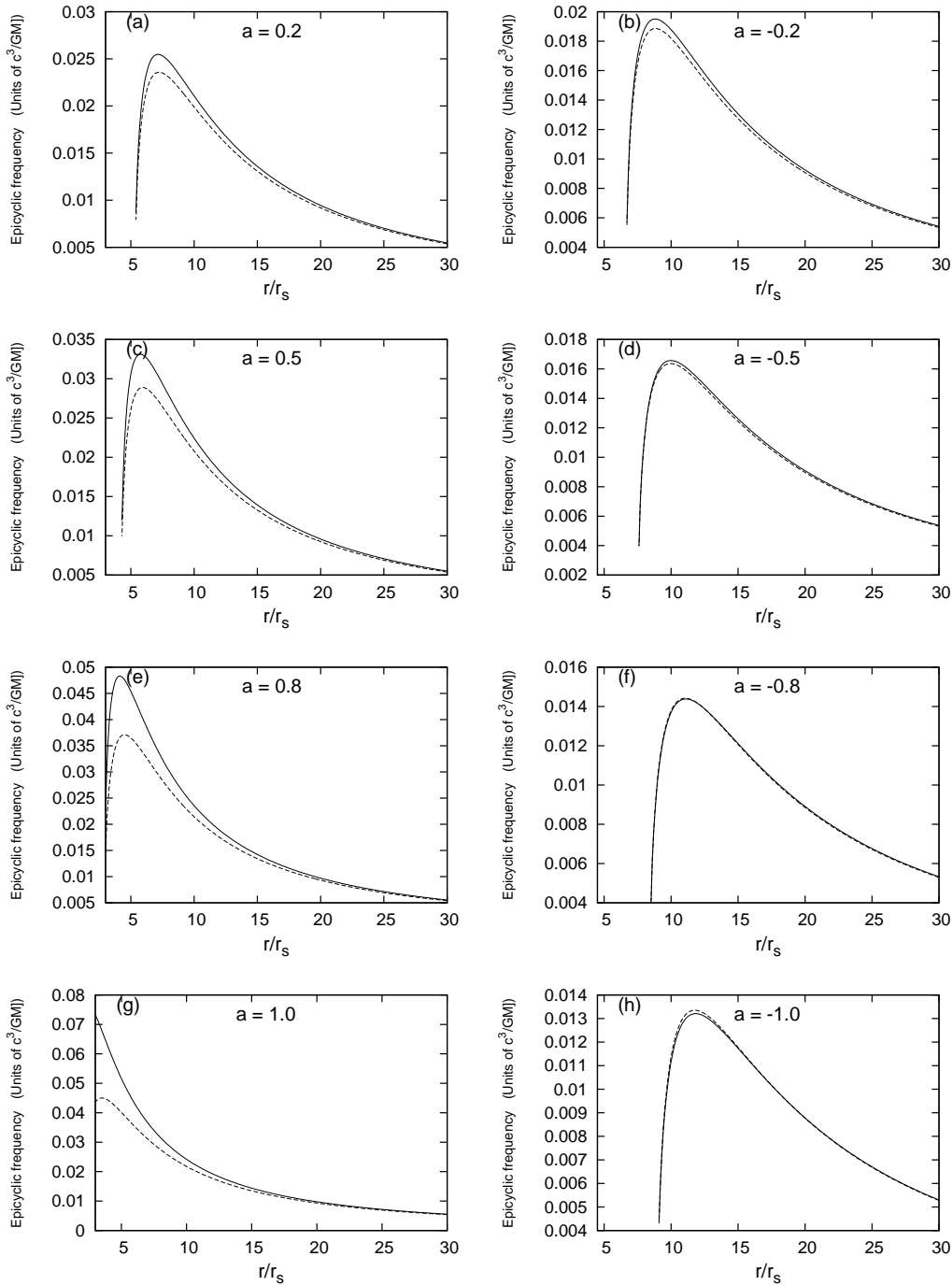}
\caption{Same as that in figures 2, 3 and 4, but variation of epicyclic frequency  
of particle in circular orbit with $r$, in units of $c^3/{GM}$.
 }
\label{Fig6}
\end{figure*}

\begin{figure*}
\centering
\includegraphics[width=170mm]{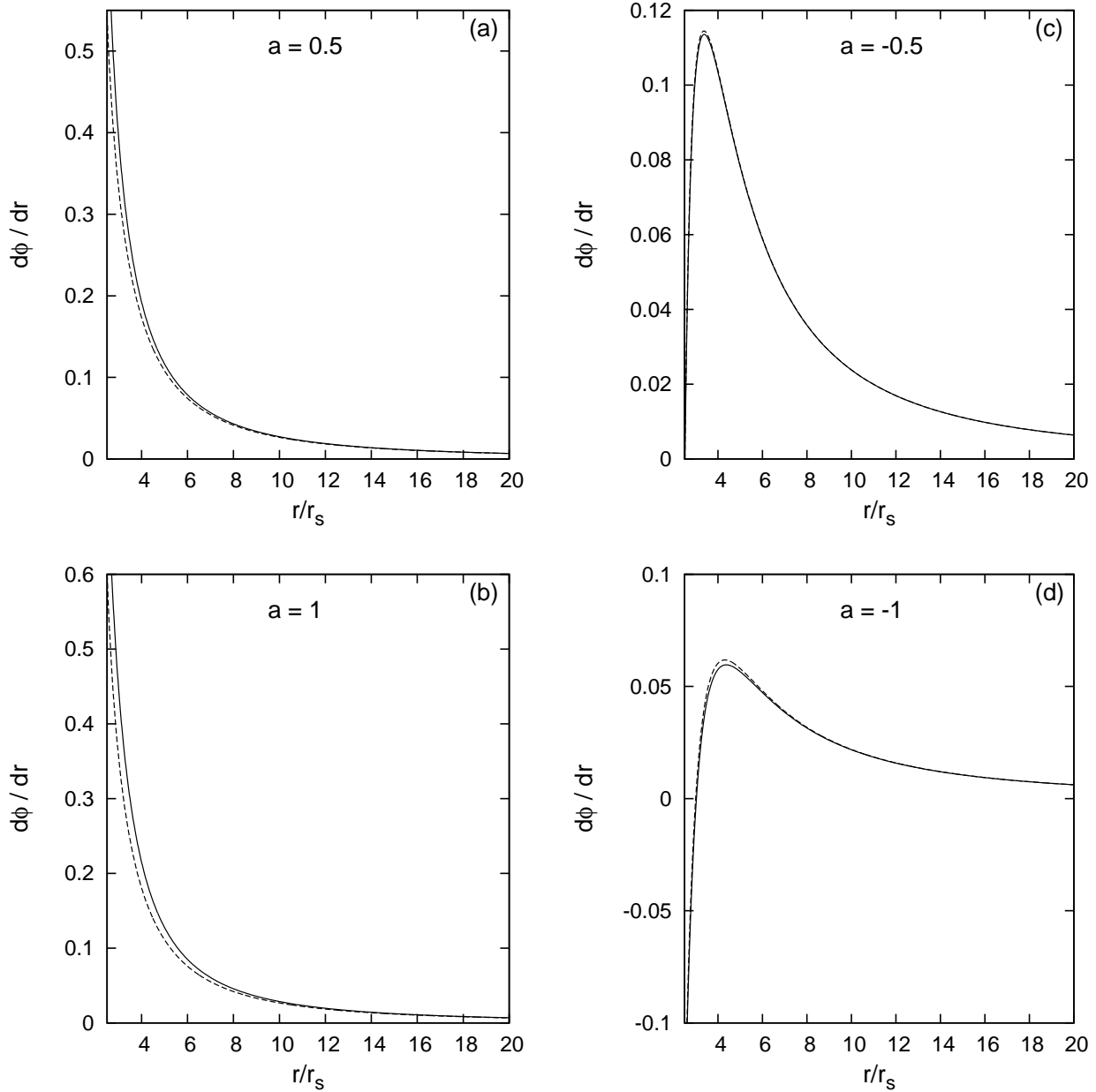}
\caption{Variation of orbital trajectory ${d\phi}/{dr}$ in radial direction $r$ for
both corotating and counterrotating particle motion with conserved specific Hamiltonian $E = 0.02$ and with specific angular momentum $\lambda = 3.5$. Solid and dashed curves are for exact Kerr geometry and Kerr-Newtonian framework respectively. 
Here energy, angular momentum and radius are in units of $c^2$, $GM/c$ and ${GM}/{c^2}$, respectively.  
}
\label{Fig7}
\end{figure*}

\begin{figure*}
\centering
\includegraphics[width=130mm]{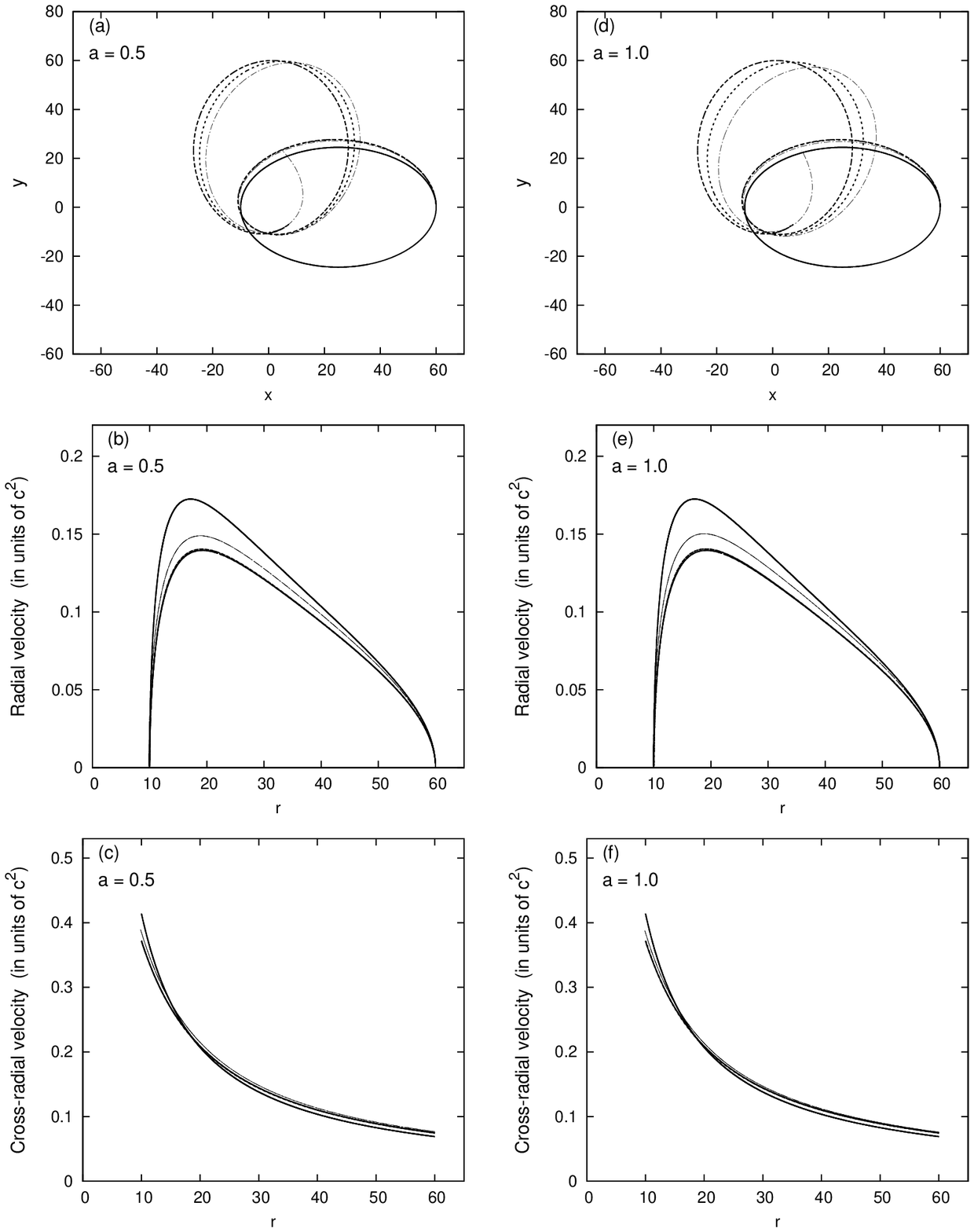}
\caption{Comparison of elliptic like trajectory of particle orbit in equatorial plane in Kerr spacetime corresponding to Kerr-Newtonian potential with that in exact general relativity, projected in x-y plane. For all cases the particle starts from an apogee $r_a = 60 \, r_s$ with eccentricity $e=0.714$.
Figures 8a,b,c correspond to $a = 0.5$. whereas figures 8d,e,f correspond to $a = 1.0$. Solid, long-dashed, short-dashed and dot-dashed curves in all the figures 
correspond to Newtonian, Schwarzschild, Kerr-Newtonian and exact Kerr geometry, respectively. The velocities are expressed in units of c.
}
\label{Fig8}
\end{figure*}

Moreover, owing to the similar nature of the orbital trajectory corresponding to Kerr-Newtonian potential and that its GR counterpart, we can conservatively predict that the Kerr-Newtonian potential would also reproduce the corresponding GR bending of light with reasonable accuracy. 

\begin{table}
\large
\centerline{\large Table 1}
\centerline{\large Comparison in the values of apsidal precession}
\centerline{\large between Kerr and Kerr-Newtonian for different $a$.}
\centerline{\large $r_a = 60 \, r_s$, $e=0.714$}
\begin{center}
\begin{tabular}{cccccccccccc}
\hline
\hline
\noalign{\vskip 2mm}
$a$ & ${\rm Kerr}$ & ${\rm Kerr-Newtonian}$ & ${\rm Error (percent)}$  \\
\hline
\hline
\noalign{\vskip 2mm} 
$0.0$  &  1.5095  & 1.5095  & 0.0 \\
\hline
\noalign{\vskip 2mm}
$0.3$  &  1.4314  & 1.4831  & 3.6118 \\
\hline
\noalign{\vskip 2mm}
$0.5$  &  1.3844 & 1.4604  & 5.4897 \\
\hline
\noalign{\vskip 2mm}
$0.8$  &  1.3276  &  1.4317  & 7.8412  \\
\hline
\noalign{\vskip 2mm}
$0.95$  & 1.2791  & 1.4145  & 10.5855  \\
\hline
\noalign{\vskip 2mm}
$1.0$  &  1.2401  & 1.3915  & 12.2087 \\
\hline
\hline
\end{tabular}
\end{center}
\end{table}

\section{A comparative analysis of the Kerr-Newtonian potential}

The essential philosophy of the procedure adopted in the present work to derive Newtonian like analogous potential of the corresponding Kerr geometry is to reproduce the geodesic equations of motion of test particles with reasonable accuracy, if not exactly. Therefore, not only does the adopted method demand that the dynamical 
profiles (such as conserved angular momentum and conserved energy) and the temporal features (such as angular and epicyclic frequencies) are reproduced wit
precise/good accuracy but also, most importantly, it guarantees the replication of the orbital trajectory of test particle motion with reasonable accuracy. In Fig. 9, we show the percentage deviation of various dynamical quantities in circular geodesics obtained with the Kerr-Newtonian potential from those of pure GR as a function of Kerr parameter $a$, at two different radii. It is to be noted that owing to the non appearance of stable circular orbits corresponding to counterrotating BHs at $r \leq 6 \, r_s$, we do not obtain any physically correct value of radial epicyclic frequencies for counterrotating circular orbits at that radii, which is being reflected in Fig. 9d. 

\begin{figure*}
\centering
\includegraphics[width=170mm]{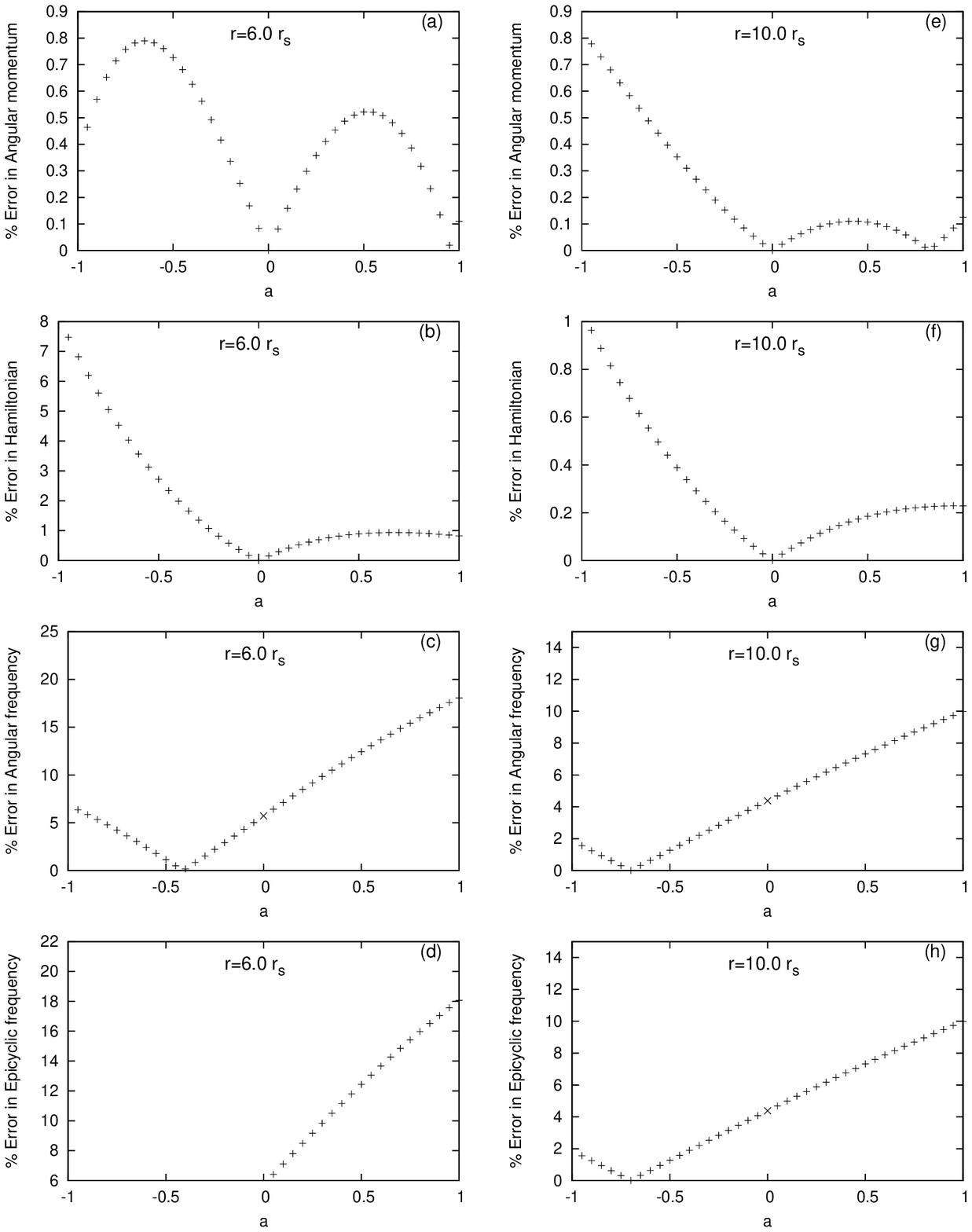}
\caption{Percentage deviation of specific angular momentum, specific Hamiltonian, specific angular frequency and specific 
epicyclic frequency corresponding to the Kerr-Newtonian potential from that of GR results, as a function of Kerr parameter $a$. Figures 10a,b,c,d correspond 
to $r =6 \, r_s$, whereas,  Figures 10e,f,g,h correspond to $r =10 \, r_s$. 
 }
\label{Fig9}
\end{figure*}

The form of the Kerr-Newtonian potential in equation (\ref{11}) is generically valid beyond $r \sim 2 r_s$. However, no real solutions for $r_{\rm ms}$ and $r_{\rm mb}$
exist beyond $a \sim 0.7$ for circular geodesics, which seems mainly due to the approximation of the low energy limit of the test particle motion used to derive Kerr-Newtonian potential. Only when restricted to the radial range $r \, \gsim \, 3 \, r_s$, the Kerr-Newtonian potential can describe circular geodesics with reasonable accuracy for any a including $a \, \gsim \, 0.7$. On the other hand, for counterrotating particle orbits, entire spectrum of GR features can be described by Kerr-Newtonian potential with precise accuracy for all values of Kerr parameter $a$. In contrast, most other prevailing PNPs corresponding to Kerr spacetime (see introduction) mostly lay emphasis to reproduce last stable circular and/or marginally bound orbits. None of them can reproduce the entire spectrum of GR features, even with marginal accuracy and within acceptable limits of error. Moreover, as the Kerr-Newtonian potential has been derived from the conserved Hamiltonian of the motion, it contains the explicit information of the velocity of the test particle as should be the case for any relativistic analogue, as well as explicit information of the frame dragging, unlike most of the PNPs corresponding to Kerr geometry. 

The PNPs corresponding to Kerr geometry are found to be less accurate than their Schwarzschild counterparts. Most of the PNPs corresponding to Kerr geometry are free fall type potentials (e.g., Artemova et al. 1996; M02; MM03), without explicit information of frame dragging. Free fall type PNPs might have some merit in mimicking spherically symmetric spacetimes, however, can be questionable while describing axially symmetric rotating BHs. Artemova et al. (1996) proposed two types of PNPs, which, however, can only reproduce the location of $r_{\rm ms}$. Moreover, it has been pointed out by M02 that the PNPs of Artemova et al. (1996) are only valid for corotating BHs. For counterrotating BHs, they furnish incorrect value of $r_{\rm ms}$. Also, the PNPs render huge error while reproducing $r_{\rm mb}$ ($\sim 500 \%$) and the specific energy in the innermost region ($\sim 50 \%$), corresponding to counterrotating BHs. Nevertheless, a few features of Keplerian accretion disk like optical depth and temperature in the GR paradigm can be reproduced by their PNPs, within an acceptable limit of error $[\sim (10-20) \%]$, for corotating BHs. They can not describe the orbital trajectories. 

Semer\'ak \& Karas (1999) prescribed a PNP, ad hocly taking into account the effect of frame dragging through a correction term. It is a three dimensional potential and has been prescribed to be useful for off-equatorial orbits. Although, the significance of geodesic equations of motion was apparently considered while prescribing their potential, however, the PNP cannot reproduce $r_{\rm ms}$ and $r_{\rm mb}$, as well as other features of circular geodesics, with reasonable accuracy. 
Semer\'ak \& Karas pointed out that the PNP is unable to approximately reproduce GR profiles of angular momentum and energy as well as the orbital trajectories, even within the acceptable limits of error. Corresponding to the Kerr-Newtonian potential, the specific energy marginally deviates from the GR results, however, with a maximum error margin of $\sim 10 \%$ in the vicinity of an extremally spinning BH, for corotating case. For counterrotating BHs, the Kerr-Newtonian potential reproduce near exact GR results for circular geodesics. 

The PNP prescribed by M02 comparatively gives far better results than the above stated PNPs, in mimicking key GR features of Kerr geometry. The PNP of M02 has been derived from the corresponding metric and it exactly reproduces $r_{\rm ms}$ for all values of $a$. Moreover, the marginally bound orbit $r_{\rm mb}$ can also be reproduced by this PNP for all values of $a$ within an error margin of $\sim 5 \%$. Also, the profile of the conserved specific energy for circular geodesics can be approximately reproduced, at least, within the acceptable limits of error (see Fig. 5 in GM07). However, the potential of M02 can not well reproduce the corresponding GR angular and epicyclic frequencies; the error margin for these parameters are as high as $\sim 180 \%$ and $\sim 800 \%$ respectively, for extremely spinning BH (MM03). Also this potential is unable to reproduce orbital trajectories properly. In MM03, two PNPs were prescribed for describing temporal effects like angular and epicyclic frequencies as well as specific energies around Kerr geometry. However, none of them can reproduce well both specific energy and angular frequencies simultaneously; for logarithmically modified potential the deviation in specific energy is more than $30 \%$ whereas the deviations in epicyclic frequencies for second-order expansion potential range from $25 \%$ to $170 \%$ when $a \le 0.9$. Moreover, these potentials can not reproduce $r_{\rm mb}$. 

A few more PNPs corresponding to the generalized Kerr geometry (three dimensional) also exist in literature. One of them is GM07 which is an extension of M02, and thus exhibits similar behavior with similar kind of limitations. Another is prescribed by Chakrabarti \& Mondal (2006) where the information of frame dragging has been ad hocly introduced. In the equatorial plane, this potential is valid approximately up to Kerr parameter $a \sim 0.8$. At this value of $a=0.8$, the value of $r_{\rm mb}$ corresponding to this PNP deviates by more that $20 \%$ from the exact GR result. Also, the dynamical profiles and the orbital trajectories are less accurately reproduced by this potential as compared to the Kerr-Newtonian. 
 
Thus, we can conclude that none of the prevailing PNPs corresponding to Kerr geometry can reproduce well all the essential GR features simultaneously, within a reasonable margin of error. In contrast within the criteria of the `low energy limit', the Kerr-Newtonian potential can approximate most of the GR features of Kerr geometry with precise/reasonable accuracy for $-1 \, \lsim \, a \, \lsim \, 0.7$. For $a > 0.7$, the circular geodesics can still be treated by Kerr-Newtonian potential accurately if one restricts to the radial range $r \, \gsim \, 3 \, r_s$. For general orbital trajectories (without confining to circular orbits only), however, the Kerr-Newtonian potential can be effectively used for $r \ge 2 \, r_s$ without any restriction on $a$. For instance, we have obtained the elliptic orbital trajectory down to $r \sim 2 r_s$ for $a=1$, using the Kerr-Newtonian potential, as shown in Fig. 10. As the Kerr-Newtonian potential describes the corresponding GR orbital trajectories with reasonable accuracy, the potential can well reproduce the experimentally tested GR effects like perihelion advancement or gravitational bending of light, within an acceptable margin of error. 

\begin{figure}
\includegraphics[width=85mm]{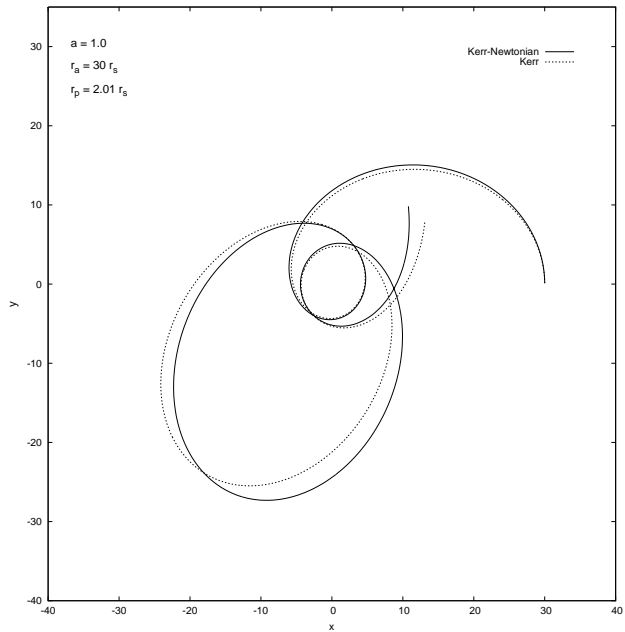}
\caption{Elliptic orbital trajectory in equatorial plane up to $r \sim 2 \, r_s$ corresponding to Kerr-Newtonian potential. Solid line is for Kerr-Newtonian and dashed line is for Kerr geometry. The particle orbit lies between $r_a = 30 \, r_s$ and $r_p = 2.01 \, r_s$ for Kerr parameter $a=1$. 
 }
\label{Fig10}
\end{figure}

\section{Discussion}

A PNP corresponding to Kerr geometry is more inconspicuous as it requires to mimic several explicit Kerr features like frame dragging and gravitomagnetic effects and therefore it is more complex. The Kerr-Newtonian potential formulated in this work invoking a physically correct methodology is found to approximate all the Kerr features with reasonable accuracy for $a<0.7$ unlike the prevailing PNPs for the Kerr space time. The formulated Kerr-Newtonian potential has been derived under low energy limit ($\epsilon/c^2 \sim 1$) approximation and also restricting the particle orbits in the equatorial plane only but even in such restricted circumstances, it is much complicated in comparison to that in pure Schwarzschild case. The analytical form of the Kerr-Newtonian potential which has been evaluated from the conserved Hamiltonian (\ref{9}) restricts its applicability to $r \, \gsim \, 2 r_s$. This would not cause any major difficulties for astrophysical scenarios as accretion studies are mainly focused on regions with $r > 2 r_s$. 

Although the robustness of Kerr-Newtonian potential in (\ref{11}), and its ability to mimic most of the GR features within an acceptable error margin is quite appreciable, however, one needs to remember that any analogous `modified Newtonian' description of relativistic geometry is inherently approximate in nature. Moreover, here we have also assumed the criteria of `low energy limit' to derive the said potential. The low energy limit criteria may be suitable to describe static geometries, for which, the results would be precisely exact and the GR features in its entirety would be reproduced with remarkable accuracy (Sarkar et al. 2014). For axially symmetric Kerr geometry, it provides more limitations. Owing to which, although the counterrotating particle orbits have been accurately described, the Kerr-Newtonian potential can not be used for $a>0.7$ in describing the innermost circular geodesics that essentially determines the gravitational energy to be extracted from matter accreting onto black hole; the potential can be employed for $a>0.7$ only when restricted to the radial range $r \, \gsim \, 3 \, r_s$. For general orbital trajectories, however, no such restriction on $a$ need to be imposed; the geodesics are permissible over the entire radial range of $r \, \gsim \, 2 \, r_{s}$, for any value of $a$. The derived potential, therefore, could be used comfortably in studying realistic astrophysical processes around rapidly spinning BHs, at least for $a<0.7$. 

The most appropriate physical system to use this kind of potential is the accreting BHs, as accretion of gaseous plasma around BHs is one of the few plausible ways to realize the presence of astrophysical BHs in the observed universe. Realistic accreting plasma dynamics is extremely complex, comprising of several microphysical 
processes. GR plasma equations with all the underlaid physical processes become extremely tedious, which then inevitably necessitates to study these systems in the Newtonian hydrodynamical/magnetohydrodynamical framework, however, with a correct Newtonian analogue of GR effects. The Newtonian framework gives us the freedom to construct more robust accretion flow models with the detailed inclusion of two temperature non-equilibrium plasma dynamics, effect of collisionless plasma, precise radiative transfer equations and other necessary turbulent diffusive terms, especially around spinning BHs. 

It is worthwhile to mention that BH accretion and related processes are also studied through GR magnetohydrodynamical (numerical) simulations (see for instance Abramowicz \& Fragile 2011 and references therein; McKinney \& Gammie 2004 ; Hawley \& Krolik 2006; Komissarov et al. 2007; McKinney et al. 2012). Recently a few full three dimensional general relativistic radiative MHD codes have been developed to study BH accretion which include COSMOS++ (Dibi et al. 2012, Fragile et al. 2012), KORAL (Sadowski et al. 2014), GRHydro (Mosta et al. 2014). The later code has been built within the framework of Einstein tool kit. Such simulation study, however, requires an expensive fast computing system and even with such facility a dynamical study can be performed at present only for a very limited time duration considering only a subset of physics. Simulating accretion disks for very large range of scales that can be present in a real system is also very difficult, if not impossible with the present day computational facility. So the PNPs are still useful to understand the underlying physics of accretion disk/jet. 

Perhaps, rotating BHs are universally present both in local universe in BHXRBs and in the center of galaxies. BH spin is directly responsible for plausibly powering astrophysical jets, generating QPOs in BHXRBs, increasing the radiative efficiency of accretion flow and several other accretion related processes. Incorporation of these effects with appropriate physics in the accreting plasma dynamics in exact GR framework renders extreme difficulty. The Kerr-Newtonian potential would then, in principle, becomes effective in studying accretion flow and its implications around rotating BHs, avoiding GR fluid equations. The real test would be to use them in real accretion scenarios in numerical and magnetohydrodynamical simulation studies. 

\section*{Acknowledgments}
The authors would like to thank an anonymous referee for insightful comments and suggestions.



\onecolumn
\noindent{\bf APPENDIX 1} \\

Here we furnish the Cartesian transformation of the acceleration terms given by equations (16) and (17) for the test particle motion in the equatorial plane, corresponding to the Kerr-Newtonian potential. Using the following identities (see TR13): \\
$$
r= \sqrt{\left(x^2 + y^2 \right)} , 
\eqno(A1)
$$
$$
\dot r = \frac{x \dot x + y \dot y}{\sqrt{\left(x^2 + y^2 \right)}}
\eqno(A2)
$$
and 
$$
\dot \phi= \frac{\left(x \dot y - y \dot x \right)}{x^2 + y^2}, 
\eqno(A3)
$$
the acceleration of the particle motion in the $x$ and $y$ directions are then given by 
$$
\ddot{x} = \frac{x}{r} \left(\ddot{r} - r {\dot \phi}^2 \right) -  \frac{y}{r} \left( r \ddot{\phi} + 2 \dot{r} {\dot \phi}  \right)
\eqno(A4)
$$
and
$$
\ddot{y} = \frac{y}{r} \left(\ddot{r} - r {\dot \phi}^2 \right) +  \frac{x}{r} \left( r \ddot{\phi} + 2 \dot{r} {\dot \phi}  \right) \, , 
\eqno(A5)
$$
where, $\ddot{r}$ and $\ddot{\phi}$ are given by equations (16) and (17), respectively. With $a=0$, the corresponding 
acceleration terms in equations (A4) and (A5) exactly reduce to those given in 
TR13. The corresponding $\ddot{r}$ and $\ddot{\phi}$ equations in Kerr geometry in the equatorial plane are given by 
$$
\ddot{r} - \frac{2}{r^2 \Delta} \left[\left(r^2 + 2 a^2 \right) r_s - \frac{a^2}{2} \left(r-2r_s \right)  - \left(3 r^2 +a^2 \right) \left(r-2 r_s \right) \frac{\omega {\dot \phi}}{2} \right] {\dot r}^2 
- \left[\frac{\Delta^2}{r^2} \frac{r-3r_s}{\left(r-2 r_s \right)^2} - \frac{a^2 \Delta}{r^3} \right] {\dot \phi}^2
$$
$$
+ \frac{GM}{r^4} \left(1+\omega \dot \phi \right)^2 \left[\frac{c^4}{\epsilon^2} 
\left(r-2 r_s \right)^2 \left(1-\frac{2a^2}{r^2} \right)  + \frac{a^2}{r} 
\left(3r-4r_s \right)  - \frac{2 \Delta r}{r-2r_s} \frac{a}{c} \frac{\dot \phi}{\left(1+\omega \dot \phi \right) }  \right]
= 0 \, 
\eqno(A6)
$$
and 
$$
\ddot{\phi} + \frac{1}{\Delta} \left[2 \left(r-3r_s \right) - \frac{4a^2 \, r_s}{r^2} \frac{r-r_s}{r-2r_s} 
+ \frac{\omega {\dot \phi} \left(3r^2+a^2  \right) \left(r-2r_s \right)}{r^2}  \right]   {\dot r} {\dot \phi} + \frac{a \, r_s \, c }{r^2 \Delta} 
\left(1+\omega \dot \phi \right)  {\dot r} = 0 \, ,
\eqno(A7)
$$
respectively. The equations (A6) and (A7) exactly reduce to that in Schwarzschild case with $a=0$ (see TR13). 
$$
$$
\noindent{\bf APPENDIX 2} \\

Here, we show the derivation of radial epicyclic frequency $\kappa$ corresponding to the Kerr-Newtonian potential. Following \S 3.2, the linearized perturbed equations are given by \\
$$
\delta {\ddot{r}}   = \delta r \, {{\dot \phi} {\vert}_C}^2 \left[1 - a^2 (r^2 - 10 r r_s + 10 r^2_s) + 8 \frac{r_s (r-r_s) a^4}{r^3 (r-2 r_s)^3} \right] 
+  \delta r  \, {{\dot \phi} {\vert}_C}^2  
\left[\frac{\Delta}{2 r (r - r_s)} -\frac{\Delta (3 r - 2 r_s) a^2}{2 r^2 (r - 2 r_s)^3} \right] \frac{\omega \, {\dot \phi} {\vert}_C}{1 + \omega \, {\dot \phi} {\vert}_C}
$$
$$
- \delta r \, \frac{{{\dot \phi} {\vert}_C}^2}{r-2r_s} 
\left[ {\mathcal D}_1  \frac{\omega \, {\dot \phi} {\vert}_C}{1 + \omega \, {\dot \phi} {\vert}_C} + {\mathcal D}_2  \frac{\omega \, {\dot \phi} {\vert}_C \, \left(1-\omega \, {\dot \phi} {\vert}_C \right)}{\left(1 + \omega \ {\dot \phi} {\vert}_C\right)^2} \right] 
+ \delta r \,\frac{2GM}{r^5} 
\left[(r-2r_s)(r-4r_s) + a^2 \left(2-\frac{5 r_s}{r} \right) \right]  
\left(1 + \omega \, {\dot \phi} {\vert}_C \right)  
$$
$$
- \,  \delta r \, \frac{4 G M r_s}{r^5} \frac{a}{c} \frac{r-r_s}{(r-2 r_s)^2}   
\left[\Delta \frac{5 r^2 - 14 r r_s + 10 r^2_s}{r(r-r_s)} - 2 (r-r_s) (r-2r_s) \right] 
{\dot \phi} {\vert}_C \, \left(1 + \omega \ {\dot \phi} {\vert}_C \right) 
+ \,\delta {\dot \phi} \frac{4 G M r_s}{{\mathcal G}_1  r^2} \frac{a}{c} \frac{r-r_s}{(r-2 r_s)^2} \left(1 + \omega \, {\dot \phi} {\vert}_C \right) 
$$
$$
- \delta {\dot \phi} \, {\dot \phi} {\vert}_C  
\left[ {\mathcal D}_1 \left(2 - 
\frac{\omega \ {\dot \phi} {\vert}_C }{1 + \omega \, {\dot \phi} {\vert}_C} \right) \right] 
+ \delta {\dot \phi} \, {\dot \phi} {\vert}_C 
 \left[{\mathcal D}_2 \left(2 \frac{\omega \, {\dot \phi} {\vert}_C }{1 + \omega \, {\dot \phi} {\vert}_C} 
+ \frac{\omega \, {\dot \phi} {\vert}_C \, \left(1-\omega \, {\dot \phi} {\vert}_C \right)}{\left(1 + \omega \, {\dot \phi} {\vert}_C\right)^2} \right) 
\right] 
\eqno(A8)
$$
and 
$$
\delta {\ddot{\phi}} = - \, \frac{\delta {\dot r}  \, {\dot \phi} {\vert}_C}{2(r-2r_s)} \left[ \omega \, {\dot \phi} {\vert}_C \left(3 + \omega \, {\dot \phi} {\vert}_C \right) \right] 
- \, \frac{\delta {\dot r}  \, {\dot \phi} {\vert}_C}{2(r-2r_s)} \frac{4 G M r_s}{{\mathcal G}_2  r^4} \frac{a}{c} \frac{r-r_s}{(r-2 r_s)^2} \left(1 + \omega \, {\dot \phi} {\vert}_C \right)^3  
$$
$$
- \, \frac{\delta {\dot r}  \, {\dot \phi} {\vert}_C}{2(r-2r_s)} \left[ \frac{1}{\Delta} 2 (r-3r_s)(r-2r_s) - 4 \frac{r_s}{r} a^2 
\right] \nonumber \\
\left(2+ \omega \, {\dot \phi} {\vert}_C \right) \left(1+ \omega \, {\dot \phi} {\vert}_C \right) \, , 
\eqno(A9)
$$
$$
$$
respectively. Here, ${\mathcal D}_1 = -\frac{{\mathcal G}_2}{2 r} \left[2 (r-2r _s)(r-3r_s) - \frac{4 r_s a^2}{r} \right]$ and ${\mathcal D}_2 = \frac{{\mathcal G}_2 \Delta}{2 r}$. Equations (A8) and (A9) reduce to that in Schwarzschild case 
with $a=0$. We assume perturbed quantities $\delta r = \delta r_0 \exp^{\imath \kappa t}$ and $\delta \phi = \delta \phi_0 \exp^{\imath \kappa t}$ for harmonic 
oscillations, where $\kappa$ is the radial epicyclic frequency. $\delta r_0$ and $\delta \phi_0$ are amplitudes and $\imath = \sqrt(-1)$ 
(Semer\'ak \& Z\'acek 2000; TR13). With the substitution of $\delta r$ and $\delta \phi$, equations (A8) and (A9) reduce to \\
$$
-\kappa^2  \, \delta r  =  {{\dot \phi} {\vert}_C}^2 \left[1 - a^2 (r^2 - 10 r r_s + 10 r^2_s) + 8 \frac{r_s (r-r_s) a^4}{r^3 (r-2 r_s)^3} \right] \, \delta r \,
+ {{\dot \phi} {\vert}_C}^2  
\left[\frac{\Delta}{2 r (r - r_s)} -\frac{\Delta (3 r - 2 r_s) a^2}{2 r^2 (r - 2 r_s)^3} \right] \frac{\omega \, {\dot \phi} {\vert}_C}{1 + \omega \, {\dot \phi} {\vert}_C} \, \delta r 
$$
$$
- \, \frac{{{\dot \phi} {\vert}_C}^2}{r-2r_s}  
\left[ {\mathcal D}_1  \frac{\omega \, {\dot \phi} {\vert}_C}{1 + \omega \, {\dot \phi} {\vert}_C} + {\mathcal D}_2  \frac{\omega \, {\dot \phi} {\vert}_C \, \left(1-\omega \, {\dot \phi} {\vert}_C \right)}{\left(1 + \omega \ {\dot \phi} {\vert}_C\right)^2} \right] \, \delta r 
+ \frac{2GM}{r^5}  
\left[(r-2r_s)(r-4r_s) + a^2 \left(2-\frac{5 r_s}{r} \right) \right] \left(1 +  \omega \, {\dot \phi} {\vert}_C \right)  \, \delta r 
$$
$$
- \,  \delta r \, \frac{4 G M r_s}{r^5} \frac{a}{c}  
\frac{r-r_s}{(r-2 r_s)^2}\left[\Delta \frac{5 r^2 - 14 r r_s + 10 r^2_s}{r(r-r_s)} - 2 (r-r_s) (r-2r_s) \right] {\dot \phi} {\vert}_C \, \left(1 + \omega \ {\dot \phi} {\vert}_C \right)  
$$
$$
+ \, \imath \kappa \, \frac{4 G M r_s}{{\mathcal G}_1  r^2} \frac{a}{c} \frac{r-r_s}{(r-2 r_s)^2} 
\left(1 + \omega \, {\dot \phi} {\vert}_C \right) \, \delta \phi  \, 
- \imath \kappa \,{\dot \phi} {\vert}_C   
\left[ {\mathcal D}_1 \left(2 - 
\frac{\omega \ {\dot \phi} {\vert}_C }{1 + \omega \, {\dot \phi} {\vert}_C} \right) 
- {\mathcal D}_2 \left(2 \frac{\omega \, {\dot \phi} {\vert}_C }{1 + \omega \, {\dot \phi} {\vert}_C} + \frac{\omega \, {\dot \phi} {\vert}_C \, \left(1-\omega \, {\dot \phi} {\vert}_C \right)}{\left(1 + \omega \, {\dot \phi} {\vert}_C\right)^2} \right) 
\right] \,    
$$
$$
\eqno(A10)
$$
and  
$$
\kappa {\delta \phi}  =  \frac{{\dot \phi} {\vert}_C}{2(r-2r_s)} \left[ \omega \, {\dot \phi} {\vert}_C \left(3 + \omega \, {\dot \phi} {\vert}_C \right)\right] \, \imath \delta r 
- \, \frac{{\dot \phi} {\vert}_C }{2(r-2r_s)} \, \frac{4 G M r_s}{{\mathcal G}_2  r^4} \frac{a}{c} \frac{r-r_s}{(r-2 r_s)^2} 
\left(1 + \omega \, {\dot \phi} {\vert}_C \right)^3 \, \imath \delta r 
$$
$$ 
+ \frac{{\dot \phi} {\vert}_C}{2(r-2r_s)} \frac{1}{\Delta} \left[2 (r-3r_s)(r-2r_s) - 4 \frac{r_s}{r} a^2 
\right] 
\left(2+ \omega \, {\dot \phi} {\vert}_C \right) \left(1+ \omega \, {\dot \phi} {\vert}_C \right) \imath \delta r  \, ,
$$
$$
\eqno(A11)
$$
respectively. Solving equations (A10) and (A11), we eventually solve for radial epicyclic frequency $\kappa$ given by equation (28) in \S 3.2. ${\mathcal F}_1$ 
and ${\mathcal F}_2$ in equation (28) are given by 
$$
{\mathcal F}_1 =  \frac{4 G M r_s}{{\mathcal G}_1  r^2} \frac{a}{c} \frac{r-r_s}{(r-2 r_s)^2} \left(1 + \omega \, {\dot \phi} {\vert}_C \right) \,
- \, {\dot \phi} {\vert}_C \, \nonumber \\ 
\left[ {\mathcal D}_1 \left(2 - 
\frac{\omega \ {\dot \phi} {\vert}_C }{1 + \omega \, {\dot \phi} {\vert}_C} \right) 
- {\mathcal D}_2 \left(2 \frac{\omega \, {\dot \phi} {\vert}_C }{1 + \omega \, {\dot \phi} {\vert}_C} + \frac{\omega \, {\dot \phi} {\vert}_C \, \left(1-\omega \, {\dot \phi} {\vert}_C \right)}{\left(1 + \omega \, {\dot \phi} {\vert}_C\right)^2} \right) \, 
\right] 
\eqno(A12)
$$
and \\
$$
{\mathcal F}_2 = \frac{{\dot \phi} {\vert}_C}{2 (r-2r_s)} \left[ \omega \, {\dot \phi} {\vert}_C \left(3 + \omega \, {\dot \phi} {\vert}_C \right) \right] 
\frac{4 G M r_s}{{\mathcal G}_2  r^4} \frac{a}{c} \frac{r-r_s}{(r-2 r_s)^2} \left(1 + \omega \, {\dot \phi} {\vert}_C \right)^3 
$$
$$
+ \frac{{\dot \phi} {\vert}_C}{2 (r-2r_s)}  
\frac{1}{\Delta} \left[2 (r-3r_s)(r-2r_s) - 4 \frac{r_s}{r} a^2 
\right] 
\left(2+ \omega \, {\dot \phi} {\vert}_C \right) \left(1+ \omega \, {\dot \phi} {\vert}_C \right) 
\frac{4 G M r_s}{{\mathcal G}_2  r^4} \frac{a}{c} \frac{r-r_s}{(r-2 r_s)^2} \left(1 + \omega \, {\dot \phi} {\vert}_C \right)^3  \, ,
\eqno(A13)
$$
$$
$$
respectively. 


\end{document}